\documentstyle{mn}
\input epsf
\newcommand{\Fs}{\,^*\! F}
\newcommand{\bA}{\bmath{A}}

\newcommand{\bJ}{\bmath{J}}
\newcommand{\bj}{\bmath{j}}
\newcommand{\cE}{\bmath{\check{E}}}
\newcommand{\cB}{\bmath{\check{B}}}

\newcommand{\br}{\bmath{r}}
\newcommand{\bv}{\bmath{v}}

\newcommand{\bm}{\bmath{m}}
\newcommand{\bp}{\bmath{p}}
\newcommand{\bn}{\bmath{n}}
\newcommand{\bP}{\bmath{P}}
\newcommand{\bB}{\bmath{B}}
\newcommand{\bE}{\bmath{E}}
\newcommand{\bH}{\bmath{H}}
\newcommand{\bD}{\bmath{D}}
\newcommand{\bL}{\bmath{L}}

\newcommand{\bS}{\bmath{S}}
\newcommand{\bOm}{\bmath{\omega}}
\newcommand{\text}[1]{\quad\mbox{#1}\quad}
\newcommand{\spr}[2]{\bmath{#1} \!\cdot\! \bmath{#2}} 
\newcommand{\vpr}[2]{\bmath{#1} \!\times\! \bmath{#2}} 
\newcommand{\vdiv}[1]{\spr{\nabla}{#1}} 
\newcommand{\vcurl}[1]{\vpr{\nabla}{#1}}

\newcommand{\oder}[2]{\frac{d #1}{d #2}}
\newcommand{\pder}[2]{\frac{\partial #1}{\partial #2}}
\newcommand{\Pd}[1]{\partial_{#1}}
\newcommand{\Od}[1]{\frac{d}{d #1}}

\title{Electrodynamics of black hole magnetospheres}
\author[S.S.Komissarov] 
{ Komissarov S.S. \\  
Department of Applied Mathematics, 
The University of Leeds, Leeds.   
E-mail: serguei@amsta.leeds.ac.uk }

\begin{document} 
\label{firstpage} 
\maketitle 

\begin{abstract} 
   The main goal of this research is to get better insights into 
   the properties of the plasma filled magnetospheres of black holes 
   by means of direct numerical simulations and, ultimately, to resolve 
   the controversy surrounding the Blandford-Znajek mechanism. 
   Driven by the need to 
   write the equations of black hole electrodynamics in the form 
   convenient for numerical applications, we constructed a new system of 
   3+1 equations which not only has more traditional form than 
   now classical 3+1 system of Thorne and Macdonald but which is also 
   more general. To deal with the magnetospheric current sheets, we also developed a  
   simple model of radiative resistivity based on the inverse Compton 
   scattering of background photons. The results of numerical simulations 
   combined with simple analytical arguments allow us to make a number of 
   important conclusions on the nature of the Blandford-Znajek mechanism. 
    We show that, just like in the Penrose mechanism and in the MHD models 
   of Punsly and Coroniti, the key role in this mechanism is played by
   the black hole ergosphere. The poloidal currents are 
   driven by the gravitationally induced electric field which cannot 
   be screened within the ergosphere by any static distribution of the 
   electric charge of locally created pair plasma. Contrary to what is 
   expected in the Membrane paradigm, the energy and angular momentum are 
   extracted not only along the magnetic field lines penetrating the event 
   horizon but along all field lines penetrating the ergosphere. 
   In dipolar magnetic configurations symmetric relative to the equatorial plane 
   the force-free approximation breaks down within the ergosphere where 
   a strong current sheet develops along the equatorial plane.  
   This current sheet supplies energy and angular momentum at infinity 
   to the surrounding force-free magnetosphere.  The Blandford-Znajek 
   monopole solution is found to be asymptotically stable and causal. 
   The so-called horizon boundary condition of Znajek is shown to 
   be a regularity condition at fast critical surface.   
\end{abstract} 
\begin{keywords}
black  hole   physics:general -- magnetic fields  -- methods:numerical
\end{keywords} 

\section{Introduction} 

As the result of great advances in observational astrophysics during
the previous several decades black holes are no longer regarded as
peculiar solutions allowed by the Einstein equations which may or may not
have anything to do with real astronomical phenomena. It is now widely
believed that black holes are common in the Universe and play a key
role in the most violent space events such as activity of galactic
nuclei and gamma-ray bursts. Enormous amounts of energy released
during such events can have two different origins. First of all this
can be the gravitational energy of matter released during 
accretion onto an already existing black hole or during the  
gravitational collapse leading to formation of a new black hole.  
On the other hand, this can be the rotational energy of a black hole itself.

The notion of rotational energy of a black hole emerged as the result of
theoretical discovery made by Penrose(1969) who found that a fraction
of the Kerr hole mass can be converted, at least in principle, into
the energy of surrounding matter or radiation (see also Cristodolou
1970). He has shown that two particles may interact in the
neighbourhood of the black hole in such a way that one of the particles
acquires negative energy and eventually disappears into the hole
whereas the other particle carries away excessive positive
energy. Unfortunately, as this was shown later, such interactions are rather
special and must be very rare under typical astrophysical conditions, 
rendering the Penrose process inefficient (Bardeen et al. 1973).
However, it was also found that the electromagnetic field can be used to
extract the rotational energy of black holes too. 

First, Goldreich and
Julian(1969) analysed the vacuum solution for a rotating neutron star with
a dipolar magnetic field aligned with the rotational axis. They argued that
the rotationally induced electric field was strong enough to 
pull charged particles from the stellar surface and, thus, fill the 
surrounding space with plasma. Using
the force-free approximation to describe the produced magnetosphere
they argued that an electromagnetically driven wind would carry away
rotational energy and angular momentum of the star.

Then, Wald(1974) found a rather interesting particular solution of 
the vacuum Maxwell equations in the Kerr spacetime.  Far away from the hole  
this solution described a uniform magnetic field aligned with the 
rotational axis of a black hole. However, near the black hole it described
a strong electric field as well. Moreover, just like in the problem of an 
aligned rotator this ``gravitationally induced'' electric field had a 
significant component along the magnetic field. 
Bisnovatyi-Kogan and Ruzmaikin (1976) argued  that, in the case of
astrophysical black holes, a rather strong magnetic field of such kind 
can be generated in their accretion discs.

Finally, Blandford \& Znajek (1977) realised that the similarity
between the vacuum solution for a Kerr black hole and the 
vacuum solution for a rotating neutron star meant the possibility of 
electromagnetically driven wind from a rotating black hole, provided 
the space around the black hole could be filled with plasma.  
Moreover, they argued that, under the typical astrophysical conditions, 
the vacuum solutions were, in fact, unstable to cascade
pair production, ensuring a plentiful supply of charged 
particles. Then
they developed a general theory of force-free steady-state
axisymmetric magnetospheres of black holes and found a perturbative
solution for a slowly rotating black hole with monopole magnetic
field. The key element of this solution was Znajek's ``boundary
condition'' \cite{Z77} imposed on the event horizon. As expected, this
solution exhibited outgoing electromagnetic fluxes of energy and
angular momentum. Moreover, the electromagnetic mechanism seemed to 
be very robust and the estimated power of the wind was high enough to 
explain the energetics of radio galaxies and quasars.

In their analysis, Blandford \& Znajek (1977) used covariant equations
of electrodynamics and operated with components of the electromagnetic
field tensor and four-potential.  Later, Thorne and Macdonald
\cite{TM,MT} developed a 3+1 approach, where the equations of black hole
electrodynamics were written in more or less traditional form in terms
of the spatial vectors of electric and magnetic field as measured by the
so-called local fiducial observer (FIDO). In this work they adopted
the system of coordinates due to Boyer and Lindquist(1967) which has a
number of useful properties but also one important drawback. Just like
the system of Schwarzschild coordinates, it is singular at the event
horizon. As the result, the event horizon appears as a peculiar inner
boundary of of physical space which required rather special treatment.

A number of authors studied the properties of electromagnetic field
near the event horizon (Hanni \& Ruffini 1973; Hajicek 1974; Znajek
1977,1978; Damour 1978) and gradually a picture emerged, according to
which the event horizon could be treated as a rotating conducting
surface with surface charges, surface currents, and a finite surface
resistivity. This perfectly suited the quest of Thorne and Mackdonald
for a new formulation of the black hole electrodynamics which would make
it look similar to the classical electrodynamics.  Surprisingly enough,
the drawback of the Boyer-Lindquist coordinates seemed to turn into an
advantage. This theory of the event horizon have made a great impact on the
current perception of the Blandford-Znajek mechanism which is now widely
associated with a mental picture of magnetic field lines originating
from the horizon and being torqued by its rotation. It stimulated the
development of The Membrane Paradigm \cite{TPM} where the event
horizon is attributed with a whole range of physical properties.
However, one has to admit that, in spite of all its attractive
simplicity and mathematical correctness, this construction is purely
artificial. 
Moreover, it is rather worrying that the emphasis put on the role of 
the event horizon makes the Blandford-Znajek mechanism 
completely alien to the Penrose mechanism, where the key role is played
by the black hole ergosphere.

The electrodynamic mechanism together with the horizon theory is now
widely accepted by the astrophysical community.  In great contrast to
this mainstream trend, Punsly and Coroniti \cite{PC,PC1} and later
Punsly (see the review in Punsly,2001) completely rejected both these
theories. They argued that the event horizon cannot be regarded as a
unipolar inductor because it is causally disconnected from the
outgoing wind.  Indeed, both the fast and the Alfv\'en waves generated at the
event horizon can propagate only inwards and cannot effect the events 
in the outer space. The apparent lack of a proper unipolar
inductor in the Blandford-Znajek solution and its reliance on Znajek's
boundary condition made Punsly and Coroniti to conclude that this
solution is nonphysical and structurally unstable.  They developed 
completely different MHD models which seemed to be based on clearer
physical ideas. In brief, they argue that gravity forces
magnetospheric plasma to rotate inside the black hole ergosphere in
the same sense as the black hole and that the magnetic field 
exhibits a similar rotation because it is "frozen" into this
plasma. 

Koide (2003) carried out MHD simulations of dipolar black hole 
magnetospheres with plasma energy-density only $1\div2$ order of magnitude 
smaller than the energy density of electromagnetic field. Although the 
numerical solution did not reached a steady state in these simulations, 
the obtained results seemed to indicate that in this regime the inertial 
effects accounted for approximately half of the extracted energy. However, 
in the black hole magnetospheres filled with plasma via pair production the 
characteristic number density of particles is given by the so-called 
Goldreich-Julian density \cite{GJ,Bes,Hir-Oka}.  Under 
the typical conditions of supermassive black holes in AGNs, 
this corresponds to the mass density of pair plasma hardly exceeding 
$10^{-13}$ of the energy density of the electromagnetic field.  
Thus, even if 
the particle density is several orders of magnitude larger than the 
Goldreich-Julian one, the inertial effects are still expected to be 
negligibly small. Given the fact that, in such a degenerate regime, 
the full system of relativistic MHD presents a real challenge for 
numerical methods (see Komissarov, 2001a, Gammie et al., 2003), it is 
difficult not to conclude that the electrodynamic approach is more suitable. 
This is one  of the reasons for the renewed interest towards the 
theory of force-free 
electrodynamics \cite{Kom02a,Bland02}. 
(This does not mean that the pair production 
is the only possible way of plasma supply in the neighbourhood of 
astrophysical black holes and that the inertial effects are always 
that small. For example, the coronas of black hole accretion 
discs are likely to be filled with dense plasma pulled from the 
disc surface.)  

The recent numerical studies of the force-free magnetospheres 
of black holes added to the controversy
surrounding the Blandford-Znajek mechanism. On one hand, they revealed
the asymptotic stability of the Blandford-Znajek monopole solution
\cite{Kom01} and, thus, raised doubts about the validity of the causality
arguments by Punsly and Coroniti.  On the other hand, the 
results for the dipolar magnetospheres questioned the virtues of the 
horizon theory as well. Indeen, they clearly indicated that the
key role in the electrodynamic mechanism is played not by the 
black hole event horizon but by its ergosphere \cite{Kom02b}.
Moreover, these simulations also revealed a certain deficiency of the
force-free approximation as a dissipative current sheet was formed in
the equatorial plane within the ergosphere.

This paper is an attempt to understand the nature of the
electrodynamic mechanism and to resolve the controversy surrounding
it.  In Sec.II we show that the 3+1 equations of the black hole
electrodynamics can be written in a more general and somewhat simpler
form than in \cite{MT,TM}. In Sec.III the basic results for the force-free
magnetospheres are re-derived using the new 3+1 formulation. In order
to handle the ergospheric current sheets, one has to go beyond the
force-free approximation and consider the resistive electrodynamics. A
model of resistivity based on the inverse Compton scattering of the
background photons is described in Sec.IV.  In Sec.V we present the
results of numerical studies of the black hole magnetospheres obtained 
within the framework of resistive electrodynamics.  The implications of 
these results for the perception of the Blandford-Znajek mechanism are
discussed in Sec.VI. The more technical information about the properties 
of light surfaces, the properties of the Kerr-Schild coordinate system, 
and the details of our numerical method is presented in the Appendix.     
 
Throughout this paper we adopt $(-+++)$ signature for the spacetime
and assume that the Greek indices range from 0 to 3 and refer to  
spacetime tensors whereas the Latin ones 
range from 1 to 3 and refer to purely spatial tensors.  
In addition, we adopt such units that the
speed of light and $4\pi$ do not appear in the equations of
electrodynamics.

\section{3+1 Electrodynamics of Black Hole } 

Following Macdonald \& Thorne (1982) we adopt the foliation
approach to the 3+1 splitting of spacetime in which    
the time coordinate $t$ parametrises a suitable filling of spacetime  
with space-like hypersurfaces described by the 3-dimensional metric tensor 
$\gamma_{ij}$. These hypersurfaces may be regarded as the 
``absolute space'' at different instances of time $t$. Below we 
describe a number of useful results for further references.     
If $\{x^i\}$ are the spatial coordinates of the absolute space then   
\begin{equation} 
  ds^2 = (\beta^2-\alpha^2) dt^2 + 2 \beta_i dx^i dt + 
         \gamma_{ij}dx^i dx^j
\label{metric} 
\end{equation}

\noindent
where $\alpha$ is called the ``lapse function'' and  
$\bbeta$ is the ``shift vector''. 
In this paper we consider only such coordinates that
 
\begin{equation} 
\Pd{t} g_{\alpha\beta} =0.        
\label{stat_metr} 
\end{equation}

\noindent
The 4-velocity of the local fiducial observer, 'FIDO',
which can be described as being at rest in the absolute space, is   

\begin{equation} 
   n_\mu = (-\alpha,0,0,0).  
\label{ncov} 
\end{equation}

\noindent
The spatial components of the projection tensor, which is used to 
construct pure spatial tensors, 
\begin{equation} 
   \gamma_{\alpha\beta} = g_{\alpha\beta} + n_\alpha n_\beta ,
\end{equation}
coincide with the components of the spatial metric $\gamma_{ij}$.  
Other useful results are 

\begin{equation} 
   n^\mu = {1 \over \alpha}(1,-\beta^i),
\label{ncon} 
\end{equation}

\begin{equation} 
   g^{t \mu} = -{1 \over \alpha} n^\mu,  
\label{gcov} 
\end{equation}

\begin{equation} 
   g = -\alpha^2 \gamma,    
\label{dets} 
\end{equation}

\noindent
where 
$$
 \beta^i=\gamma^{ij}\beta_j  , \quad  
 g = \det{g_{\mu\nu}}, \quad 
 \gamma = \det{\gamma_{ij}}.  
$$
$\beta^i$ are the components of the velocity of the spatial grid relative  
to the local FIDO as measured using the coordinate time $t$ and the
spatial basis $\{ \Pd{i}\}$ \cite{MT}.

Landau and Lifshitz (1951) proposed a different approach to the 3+1
splitting of spacetime which is known as the congruence approach. They
also showed that within this approach the covariant equations of
electrodynamics in gravitational field can be reduced to a set of 3+1
equations very similar to the usual Maxwell equations in matter
(e.g. Landau and Lifshitz, 1971).

The 3+1 electrodynamics of black holes within the foliation approach 
was developed by
Thorne and Macdonald (1982) who used as a starting point the results
by Ellis(1973) on the 3+1 electrodynamics in the congruence language.  
In fact, they derived rather general 3+1 integral equations of electrodynamics 
as well as the differential equations adapted to a particular foliation 
of the Kerr space-time, namely the Boyer-Lindquist foliation.  This
formulation has been used successfully in various theoretical studies
of black hole electrodynamics. However, the Boyer-Lindquist foliation
has one important disadvantage -- it leads to the well known
coordinate singularity at the event horizon. This singularity is
removed in a different less known foliation, namely the Kerr-Schild
foliation, which otherwise shares many common properties with the
Boyer-Lindquist foliation (see Appendix B).  It has been shown that 
the Kerr-Schild
coordinates do not only simplify theoretical analysis but also allow
to overcome a number of problems concerning numerical simulations
(e.g. Papadopoulos \& Font,1998, Komissarov, 2001).  The corresponding
set of 3+1 differential equations of electrodynamics is somewhat 
different. Moreover, in
order to construct a Godunov type numerical scheme one needs 
a different type of integral equations compared to those given 
in Thorne and Macdonald (1982).  All these equations can be obtained from the
3+1 integral equations of \cite{MT}. However, they can also be 
derived directly from the covariant Maxwell equations, and so easily,  
that this is worth to be
shown (see also Landau \& Lifshitz, 1971; Koide, 2003).

The covariant Maxwell equations are  
(e.g. Jackson(1975)):

\begin{equation} 
  \nabla_\beta  \Fs^{\alpha \beta} = 0, 
\label{Maxw1}
\end{equation}

\begin{equation} 
  \nabla_\beta  F^{\alpha \beta} =  I^\alpha, 
\label{Maxw2}
\end{equation}

\noindent
where $F^{\alpha\beta}$ is the Maxwell tensor of the electromagnetic 
field, $\Fs^{\alpha\beta}$ is the Faraday tensor and  
$I^\alpha$ is the 4-vector of the electric current. 
The most direct way of 3+1 splitting of the covariant Maxwell equations 
is to write them down in components and then to introduce such spatial 
vectors that these equations have a particularly simple and familiar
form. For example, when eq.(\ref{Maxw1}) is written in components it splits 
into two parts: 
\begin{itemize} 
\item{The time part:} 

\begin{equation} 
   \frac{1}{\sqrt{\gamma}} \Pd{i} 
   \left( \alpha\sqrt{\gamma} \Fs^{t i} \right) =0,       
\label{e1}
\end{equation}

\item{The spatial part:} 

\begin{equation}
   \frac{1}{\sqrt{\gamma}} \Pd{t} 
   \left( \alpha\sqrt{\gamma} \Fs^{j t} \right) +       
   \frac{1}{\sqrt{\gamma}} \Pd{i} 
   \left( \alpha\sqrt{\gamma} \Fs^{j i} \right) =0,       
\label{e2}
\end{equation}

\end{itemize} 

\noindent
If we now introduce the spatial vectors $\bB$ and $\bE$ via 

\begin{equation} 
     B^i=\alpha \Fs^{it}
\label{B1}
\end{equation}

\begin{equation}
   E^i=\gamma^{ij}E_j, \quad  
   E_i =\frac{\alpha}{2} e_{ijk} \Fs^{jk},
\label{E1}
\end{equation}

\noindent
where $e_{ijk} = \sqrt{\gamma} \epsilon_{ijk} $ 
is the Levi-Civita pseudo-tensor of the absolute space, then
equations (\ref{e1},\ref{e2}) read 

\begin{equation} 
   \vdiv{B}=0, 
\label{divB} 
\end{equation}

\begin{equation} 
   \Pd{t}\bB + \vcurl{E} = 0,   
\label{Faraday}
\end{equation}

\noindent
where $\nabla$ is the covariant derivative of the absolute space.  
Notice, that in order to derive eq.(\ref{Faraday}) we used the stationarity 
condition (\ref{stat_metr}). 
Similarly, equation (\ref{Maxw2}) splits into 

\begin{equation} 
\label{divD} 
   \vdiv{D}=\rho, 
\end{equation}

\begin{equation} 
   -\Pd{t}\bD + \vcurl{H} = \bJ 
\label{Ampere}
\end{equation}

\noindent 
where 

\begin{equation} 
     D^i=\alpha F^{ti},
\label{D1}
\end{equation}

\begin{equation}
     H^i=\gamma^{ij}H_j, \quad H_i =\frac{\alpha}{2} e_{ijk} F^{jk}.
\label{H1}
\end{equation}

\begin{equation}
    \rho=\alpha I^t, \quad J^k=\alpha I^k. 
\label{rhoJ}
\end{equation}
As one can see, these 3+1 equations have exactly the same form 
as the classical Maxwell equations for the electromagnetic field in 
matter. Applying $\nabla$ to eq.(\ref{Ampere}) and then
using eq.(\ref{divD}) one obtains the electric charge conservation law

\begin{equation}
    \Pd{t}\rho + \vdiv{J}= 0. 
\label{ECC}
\end{equation}

Similar to any highly ionized plasma, the pair plasma of black hole 
magnetospheres has essentially zero electric and magnetic 
susceptibilities. In such a case, the Faraday tensor is simply 
dual to the Maxwell tensor   

\begin{equation} 
\Fs^{\alpha  \beta} = \frac{1}{2} e^{\alpha \beta \mu \nu} F_{\mu \nu}  
\label{dual_F}
\end{equation}

\begin{equation} 
F^{\alpha  \beta} = -\frac{1}{2} e^{\alpha \beta \mu \nu} \Fs_{\mu \nu}.   
\label{F}
\end{equation}
Here 

\begin{equation} 
e_{\alpha \beta \mu \nu} = \sqrt{-g}\,\epsilon_{\alpha \beta \mu \nu}, 
\label{LCt}
\end{equation}

\noindent
is the Levi-Civita alternating pseudo-tensor of spacetime 
and  $\epsilon_{\alpha \beta \mu \nu}$ is the four-dimensional Levi-Civita 
symbol. This allows to obtain the following alternative expressions for 
$\bB,\bE,\bD$, and $\bH$: 
\begin{equation} 
     B^i= \frac{1}{2}e^{ijk} F_{jk}
\end{equation}

\begin{equation}
   E_i = F_{it},
\end{equation}

\begin{equation} 
     D^i = \frac{1}{2} e^{ijk} \Fs_{jk},
\end{equation}

\begin{equation}
     H_i = \Fs_{ti}. 
\end{equation}

\noindent
Moreover, from the above definitions one immediately finds the 
following vacuum constitutive equations: 

\begin{equation} 
     \bE = \alpha \bD + \vpr{\bbeta}{B}, 
\label{E3}
\end{equation}

\begin{equation} 
     \bH = \alpha \bB - \vpr{\bbeta}{D}. 
\label{H3}
\end{equation}
At infinity, provided the coordinate system becomes Lorentzian 
there, one has $\alpha=1$, $\beta=0$ and, hence, 
$$
   \bB=\bH, \qquad \bE=\bD. 
$$   

\noindent
One can easily obtain the following covariant forms of  
definitions (\ref{B1},\ref{E1},\ref{D1},\ref{H1},\ref{rhoJ}): 

\begin{equation} 
     B^\mu = -\Fs^{\mu\nu} n_\nu,
\label{B2}
\end{equation}

\begin{equation} 
     E^\mu = -\frac{1}{2}\gamma^{\mu\nu} e_{\nu\alpha\beta\gamma} 
          k^{\alpha} \Fs^{\beta\gamma},  
\label{E2}
\end{equation}

\begin{equation} 
     D^\mu = F^{\mu\nu} n_\nu,
\label{D2}
\end{equation}

\begin{equation} 
     H^\mu = -\frac{1}{2}\gamma^{\mu\nu} e_{\nu\alpha\beta\gamma} 
             k^{\alpha} F^{\beta\gamma},  
\label{H2}
\end{equation}

\begin{equation} 
     J^\mu = 2I^{[\nu}k^{\mu]}n_\nu, 
\label{J2}
\end{equation}

\begin{equation}
    \rho=-I^\nu n_\nu,
\label{rho2}
\end{equation}

\noindent
where $ k^\alpha = \Pd{t}$. 
All these spacetime vectors are purely spatial, that is they have zero 
time component.  

In terms of four-potential ${\cal U}_\mu$ one has 
$$
   F_{\mu\nu}=-2 {\cal U}_{[\mu,\nu]} .
$$
Introducing the scalar potential $\Phi$ and the vector potential $\bA$ as 

\begin{equation} 
   \Phi=-{\cal U}_t , 
  \label{Phi}
\end{equation}

\begin{equation} 
   A^i=\gamma^{ij} A_i, \quad A_i={\cal U}_i  
  \label{A}
\end{equation}
one obtains the familiar looking results

\begin{equation} 
   \bE=-\nabla\Phi-\Pd{t}\bA, 
  \label{DinA}
\end{equation}

\begin{equation} 
  \bB=\vcurl{A} .
  \label{BinA}
\end{equation}
The covariant definitions are 
$$
   A^\mu=\gamma^{\mu\nu}{\cal U}_\nu, \quad 
   \Phi=-{\cal U}_\mu k^\mu. 
$$

From the covariant equation of motion for a particle with 
the four-velocity $u^\nu$, the four-momentum $p^\mu$, and 
the electric charge $q$,  
\begin{equation}
\frac{Dp_\mu}{D\tau} = q F_{\mu\nu}u^\nu , 
\end{equation}
\noindent 
one easily obtains the following energy equation  
\begin{equation}
   \frac{d e_p}{dt} = q \spr{E}{v},
\end{equation}
where $e_p=-p_t$ is known as the ``energy at infinity'' and 
$v^i=dx^i/dt$ is the spatial velocity vector. 

Both in the Boyer-Lindquist and the Kerr-Schild coordinates,  
where $\Pd{\phi}g_{\nu\mu}=0$, one has an equally simple 
result for the angular momentum at infinity, $l_p=p_\phi$,
   
\begin{equation}
   \frac{d l_p}{dt} = q (\bE + \vpr{v}{B})\!\cdot\!\bm ,
\end{equation}
where $\bm=\Pd{\phi}$.

In terms of Maxwell tensor the electromagnetic stress-energy-momentum
tensor is 

\begin{equation} 
     T^\mu_{\ \nu} = F^{\mu\gamma}F_{\nu\gamma} - 
           \frac{1}{4} (F_{\gamma\beta}F^{\gamma\beta})\delta^\mu_{\ \nu}. 
  \label{T}
\end{equation}
In terms of $\bB$,$\bE$,$\bH$, and $\bD$ we have 

\begin{equation} 
     T^t_{\ t} = -\frac{1}{2\alpha}(\spr{E}{D}+\spr{B}{H}),
  \label{T_tt}
\end{equation}
\begin{equation} 
     T^i_{\ t} = -\frac{1}{\alpha}e^{ijk}E_j H_k,
  \label{T_it}
\end{equation}
\begin{equation} 
     T^t_{\ i} = \frac{1}{\alpha}e_{ijk}D^j B^k
  \label{T_ti}
\end{equation}
\begin{equation} 
     T^i_{\ j} = -\frac{1}{\alpha}(D^iE_j + B^iH_j) + 
              \frac{1}{2\alpha}(\spr{E}{D}+\spr{B}{H})\delta^i_j.
  \label{T_ij}
\end{equation}
The covariant energy-momentum equation is

\begin{equation} 
     \nabla_\nu T^\nu_{\ \mu} = -F_{\mu\gamma} I^\gamma. 
  \label{EMC}
\end{equation}
Both in the Boyer-Lindquist and in the Kerr-Schild coordinates, 
as in any other coordinates with cyclic coordinate $\phi$ such that 
$\Pd{t} g_{\mu\nu} = \Pd{\phi} g_{\mu\nu} = 0 $, 3+1 splitting of 
eq.(\ref{EMC}) leads to the following energy 

\begin{equation} 
\Pd{t} e + \vdiv{S} = -(\spr{E}{J}),
  \label{EC}
\end{equation}
and the angular momentum 

\begin{equation} 
\Pd{t} l + \vdiv{L} = -(\rho \bE + \vpr{J}{B})\!\cdot\!\bm ,
  \label{AMC}
\end{equation} 
equations for the electromagnetic field. Here 
\begin{equation} 
  e=-\alpha T^t_{\ t} = \frac{1}{2}(\spr{E}{D}+\spr{B}{H})
  \label{e}
\end{equation} 
is the volume density of energy at infinity,  
\begin{equation} 
  l=\alpha T^t_{\ \phi} = (\vpr{D}{B})\!\cdot\!\bm 
  \label{l}
\end{equation} 
is the volume density of angular momentum at infinity, 

\begin{equation} 
      \bS=\vpr{E}{H} 
  \label{S}
\end{equation} 
is the flux of energy at infinity, and 

\begin{equation} 
      \bL=-(\spr{E}{m})\bD -(\spr{H}{m})\bB +  
              \frac{1}{2}(\spr{E}{D}+\spr{B}{H}) \bm
  \label{L}
\end{equation} 
is the flux of angular momentum at infinity.   

It is well known that evolution equations like (\ref{Faraday}) and 
(\ref{Ampere}) can be written as conservation laws. In our case we 
obtain 

\begin{equation} 
  \Pd{t} B^i + \nabla_j K^{ij} = 0, 
\label{Far1}
\end{equation}
and 

\begin{equation} 
  \Pd{t} D^i + \nabla_j L^{ij} = -J^i ,   
\label{Amp1}
\end{equation}

\noindent 
where 

\begin{equation} 
 K^{ij} = e^{ijk} E_k \text{and}   
 L^{ij} = -e^{ijk} H_k. 
\label{KL}
\end{equation}

\noindent 
are the magnetic and the electric field flux tensors. The corresponding 
integral equations are now easily obtained using the divergence theorem

\begin{equation} 
  \Od{t} \int_V B^i dV  + \int_{\delta V} K^{ij} dS_j = 0, 
\label{Far2}
\end{equation}

\begin{equation} 
  \Od{t} \int_V D^i dV + \int_{\delta V} L^{ij} dS_j = 
          -\int_V J^i dV,   
\label{Amp2}
\end{equation}
where $\delta V$ is the closed boundary of the spatial volume $V$; 
$dV=\sqrt{\gamma}dx^1 dx^2 dx^3$ is the infinitesimal metric 
volume of the absolute space, and $dS_i = e_{ijk}dx^j_{(1)} dx^k_{(2)}$ 
is the infinitesimal metric surface element of $\delta V$. 
One can see that both the differential and the integral 3+1 equations of 
black hole electrodynamics are, in fact, very similar to the
corresponding equations of Minkowski spacetime. This suggests 
that many well known techniques can, almost readily, be used 
to solve this equations numerically. 

In \cite{MT,TPM} the 3+1 equations of electrodynamics are 
written in terms of the electric and the magnetic field vectors as 
measured by the local FIDO. We denote this vectors as $\cE$ 
and $\cB$. From definitions (\ref{D2},\ref{B2}) it follows 
that 
\begin{equation} 
    \cB = \bB \text{and} \cE = \bD.      
\end{equation}      

The main benefit gained by the introduction of FIDOs and, hence, 
the quantities measured by these observers, comes from the 
principle of equivalence. These observers can be treated as 
locally inertial observers and in their frames all local physical 
phenomena are governed by the laws of special relativity. For 
example, in order to close the system of Maxwell equations we 
need an additional constitutive equation relating the electric current to
the electromagnetic field, namely the Ohm law. Thanks to the principle of
equivalence, this can be done in terms of quantities measured by FIDOs, 
in the same way as in the special relativistic electrodynamics.
From equation (\ref{rho2}) it follows that $\rho$ is, in fact, the
electric charge density as measured by FIDOs. 
As shown in \cite{TM} the electric current density as measured by 
FIDOs, $\bj$, is related to $\bJ$ via   

\begin{equation} 
     \bJ=\alpha\bj-\rho\bbeta. 
\label{e5}
\end{equation}
The second term in this equation accounts for the motion of spatial 
grid relative to FIDO.

In the Boyer-Lindquist coordinates $\spr{\nabla}{\beta}=0 $, and under this
condition equations (\ref{divB},\ref{Faraday},\ref{divD},\ref{Ampere}) 
reduce to the corresponding equations in \cite{MT,TPM}. For example, 
using equations (\ref{E3},\ref{divB}) one finds that  

\begin{equation} 
   \vcurl{\bE} = \vcurl{\alpha \cE} - {\cal L}_\bbeta\cB,  
\label{e3}
\end{equation}
where 

\begin{equation} 
  {\cal L}_\bbeta \cB = (\spr{\bbeta}{\nabla})\cB - 
                        (\spr{\cB}{\nabla})\bbeta 
\nonumber
\end{equation}
is the Lie derivative of $\cB$ along $\bbeta$. Thus, equation (\ref{Faraday})
reads 

\begin{equation} 
   \Pd{t}\cB -{\cal L}_\bbeta \cB + \vcurl{\alpha \cE} = 0,  
\label{e4}
\end{equation}
which is equation (3.52) in Thorne et al.(1986). However, 
in other coordinate systems, e.g. the Kerr-Schild system, 
$\spr{\nabla}{\beta}\not=0$.

The results presented in this section show that the 3+1 equations of black
hole electrodynamics can be written in a more general and much simpler
form than in \cite{TM,MT}. Our equations are remarkably similar to the
familiar equations of electrodynamics in matter. All effects of
gravity are hidden in the constitutive equations {\ref{E3},\ref{H3}) and in the
spatial metric $\gamma_{ij}$.

\section{Steady-state force-free 
magnetospheres of black holes}  

In this section we re-derive, within our 3+1 framework, 
some of the well known results for the 
steady-state force-free magnetospheres of black holes 
obtained by Blandford \& Znajek(1977). 
The only condition imposed on the coordinate 
system is $\Pd{\phi}g_{\nu\mu} =\Pd{t} g_{\mu\nu}=0$ and, thus, 
these results hold equally well both in the Boyer-Lindquist and the 
Kerr-Schild coordinates.     

The covariant force-free condition 
$$
   F_{\mu\nu} I^{\mu}=0  
$$ 
can be written in the 3+1 language as 

\begin{equation}
   \spr{E}{J}=0,
\label{a2} 
\end{equation} 
and 

\begin{equation}
  \rho \bE + \vpr{J}{B} =0,
\label{a3} 
\end{equation} 
or 

\begin{equation}
   \spr{\bD}{j}=0,
\label{Ej} 
\end{equation} 
and 

\begin{equation}
  \rho \bD + \vpr{j}{B} =0.
\label{EjB} 
\end{equation} 
From eq.(\ref{EjB}) it follows that 

\begin{equation}
   \spr{D}{B}=\spr{\cE}{\cB}=0,
\label{EB} 
\end{equation} 
and 

\begin{equation}
   \bj_{\perp}=\rho \frac{\vpr{D}{B}}{B^2} =    
   \rho \frac{\vpr{\cE}{\cB}}{\check{B}^2},     
\label{a10} 
\end{equation} 
where $\bj_\perp$ is the component of electric current normal to $\bB$. 
Equation (\ref{a10}) states that this component is entirely due to 
the drift motion of charged particles. Since the drift velocity must 
be lower than the speed of light we have 

\begin{equation}
   B^2-D^2 >0 \text{or}
   \check{B}^2-\check{E}^2 >0 .
\label{a11} 
\end{equation} 
Equations (\ref{EB}) and (\ref{a11}) are the 3+1 representations of the Lorentz 
invariant constraints of force-free approximation 
$$
  \Fs_{\mu\nu}F^{\mu\nu}=0 \text{and} 
  F_{\mu\nu}F^{\mu\nu}>0 
$$
\cite{Z77,Kom02a}. Whereas the constraint (\ref{EB}) is automatically 
satisfied by any force-free steady-state solution, the constraint (\ref{a11}) 
is not and, hence, it always has to be checked \cite{Z77}.  
If this condition is not satisfied the Alfv\'en wavespeed 
becomes complex and, thus, the system of force-free electrodynamics 
(magnetodynamics seems to be a better name) is no longer hyperbolic \cite{Kom02a}.
The physical reason why the force-free approximation breaks down if $\spr{E}{B}=0$
but $B^2-E^2<0$ is quite obvious. Under such conditions, one can find a 
local inertial frame where the electromagnetic field is seen as a pure 
electric field. 

The conditions of axisymmetry and time-independency imply that 

\begin{equation}
    E_\phi = {\cal U}_{t,\phi}-{\cal U}_{\phi,t} =0. 
\label{a1} 
\end{equation} 
Given this, equation (\ref{a3}) ensures that 

\begin{equation}
  \bJ_p \parallel \bB_p, \quad \bE_p \perp \bB_p, 
\label{a4} 
\end{equation} 
where index $p$ refers to the poloidal component of a vector. These 
show that there exists a purely azimuthal vector 
$\bmath{\omega} =\Omega \Pd{\phi}$ such that 

\begin{equation}
  \bE = -\vpr{\omega}{B} 
\label{a5} 
\end{equation} 
and 

\begin{equation}
  \bD = -\frac{1}{\alpha}(\bomega+\bbeta)\!\times\! \bB. 
\label{a5b} 
\end{equation} 
Substituting (\ref{a5}) into the stationary version of (\ref{Faraday}),  
one finds 
$$
   \vcurl{(\vpr{\omega}{B})} = 0, 
$$
which leads to 

\begin{equation}
  \bB\cdot\nabla\Omega=0. 
\label{a6} 
\end{equation} 
Thus, $\Omega$, which is called the angular velocity of magnetic 
field, is constant along the magnetic field lines. 

From equations (\ref{Ampere}) and (\ref{a4}) it immediately follows
that  

\begin{equation}
   B^\theta \Pd{\theta} H_\phi + B^r \Pd{r} H_\phi =0,   
\label{a7} 
\end{equation} 
and, thus, $H_\phi$ is also constant along the magnetic field lines 
(This constant is denoted as $B_T$ in Blandford \& Znajek,1977.) 

Another parameter constant along the magnetic field lines is the scalar 
potential, which we denote as $\Phi$. 
Indeed, from eqs.(\ref{DinA},\ref{a5}) one immediately finds that 
\begin{equation}
    \spr{B}{\nabla} \Phi = 0.   
\label{a8} 
\end{equation} 

Since the source terms in eqs.(\ref{EC},\ref{AMC})
vanish,  the energy and the angular momentum of the electromagnetic field 
are conserved. 
\begin{equation} 
\Pd{t} e + \vdiv{S} = 0, 
  \label{EC1}
\end{equation}

\begin{equation} 
\Pd{t} l + \vdiv{L} = 0.
  \label{AMC1}
\end{equation} 
Substituting $\bE$  from (\ref{a5})  into eq.(\ref{L}) and eq.(\ref{S}) one 
obtains the following expressions for the poloidal component of 
the angular momentum flux vector 

\begin{equation}
   \bL_p = -H_\phi \bB_p, 
\label{a13} 
\end{equation} 
and the poloidal component of the energy flux vector

\begin{equation}
   \bS_p = -(H_\phi \Omega) \bB_p, 
\label{a14} 
\end{equation} 
where $\bB_p$ is the poloidal component of the magnetic field.  Thus, both 
the energy and the angular momentum are transported along the poloidal 
field lines. 
  
Finally, from eqs.(\ref{E3},\ref{a5}) one obtains the following useful result

\begin{equation} 
   \begin{array}{ll}
   (B^2-D^2)\alpha^2 =  B^2({\alpha^2}-{\beta^2}) 
             & +( \spr{\omega}{B} +\spr{\beta}{B} )^2 \\ 
             & - B^2\left( 
                     \omega^2 +2\spr{\omega}{\beta} \right) . 
   \end{array}
\label{a12} 
\end{equation}

\section{Generalized Ohm's law}        

It is generally accepted that the magnetospheres of black holes are 
filled with perfectly conducting $e^+$-$e^-$ plasma 
\cite{BZ,Phi,Bes,Hir-Oka}. Should, however, a current sheet be 
formed in the magnetosphere (see Komissarov, 2002) the perfect 
conductivity approximation would fail and a model of electric  
resistivity would be required. It is well known that magnetic field 
strongly effects the electric properties of plasma by suppressing 
electric conductivity across the magnetic field lines and, thus, 
making the simple scalar Ohm's law inadequate. In fact, the total 
electric current splits into three components
 
\begin{equation} 
  \bj = \sigma_\parallel \cE_\parallel +    
        \sigma_\perp \cE_\perp + \bj_d, 
\label{Ohm}    
\end{equation} 
where $\cE_\parallel$ is the component of electric field parallel to
the magnetic field, $\cE_\perp$ is the component of electric field
perpendicular to the magnetic field, and $\bj_d$ is the drift current
which is perpendicular both to the electric and to the magnetic field
vectors \cite{Cow,Mestel}.  
The magnetospheric plasma is collisionless and its
resistivity arises via collective processes, ``the anomalous
resistivity'', and  emission of photons,  ``the radiative resistivity''.  
The radiative resistivity associated with the inverse Compton scattering 
of the photons emitted by the accretion disc has already been
considered as a source of the internal resistivity in the potential 
gaps of the black hole magnetospheres \cite{Bes,Hir-Oka}. 
Since, this is a rather simple and yet robust mechanism, it deserves to 
be considered in details. 

If the gyration and the collisional time-scales are much smaller
compared to the macroscopic time-scale then the mean velocity of a
charged particle in electromagnetic and radiation fields, as observed
by the local FIDO, is governed by

\begin{equation} 
  q (\cE + \vpr{v}{\cB}) - 
                 (W^2-1) \sigma_T u_b \frac{\bv}{v}=0,   
\label{motion1}   
\end{equation} 
where the last term describes the inverse Compton interaction with soft
background photons \cite{Bes,Hir-Oka}. Here $\bp$, $\bv$, $W$,
and $q=\pm e$ are the momentum, the velocity, the Lorentz factor, and
the electric charge of the particle respectively, $\sigma_T$ is the
Thomson cross section, and $u_b$ is the energy density of the
radiation field. For simplicity, we assume that the radiation field is
isotropic (Since the Boyer-Lindquist FIDO becomes singular near the
event horizon this condition cannot be satisfied there at any length. 
However, for the Kerr-Schild FIDO this seems to be more or less
acceptable everywhere.)  One can solve this equation for the
perpendicular and the parallel to the magnetic field components of
velocity to obtain

\begin{equation} 
    \bv_\parallel = \pm \frac{e}{\chi} \cE_\parallel, 
\label{v-pa}   
\end{equation} 

\begin{equation} 
    \bv_\perp = \pm\frac{e}{\chi} \frac{1}{1+\delta^2} \cE_\perp + 
         \frac{\delta^2}{1+\delta^2} \frac{\vpr{\cE}{\cB}}{\check{B}^2} ,  
\label{v-pe}   
\end{equation} 
where 
\begin{equation} 
  \chi = m W \nu_c, 
\end{equation} 

\begin{equation} 
  \delta = \nu_b/\nu_c, 
\end{equation} 
where 

\begin{equation} 
   \nu_b = \frac{e\check{B}}{m W c} 
\end{equation} 
is the electron gyration frequency, 

\begin{equation} 
   \nu_c = \frac{(W^2-1) \sigma_T u_b}{m W v} 
\end{equation} 
is the effective frequency of the inverse Compton collisions; here  
$m$ is the particle rest mass. 
The last term in eq.(\ref{v-pe}) does not depend on the sign of the 
electric charge and describes the usual drift motion effected by  
the collisions.

Equations (\ref{v-pa},\ref{v-pe}) 
give us the generalized Ohm law of the same form as
eq.(\ref{Ohm}) with 
  
\begin{equation} 
    \sigma_\parallel = \frac{n e^2}{ m W \nu_c} 
\label{sigma-pa} 
\end{equation}

\begin{equation} 
    \sigma_\perp = \frac{1}{1+\delta^2} \sigma_\parallel  
\label{sigma-pe} 
\end{equation}
  
\begin{equation} 
    \bj_d =  \rho \frac{\delta^2}{1+\delta^2} 
             \frac{\vpr{\cE}{\cB}}{\check{B}^2},  
\label{jdrift1} 
\end{equation}
and 
 
\begin{equation} 
    \rho = e (n_+ - n_-),\quad n=n_+ + n_-,  
\end{equation}
where $n_\pm$ are the number densities of electrons and positrons.

\begin{figure}
\leavevmode
\epsffile[0 20 190 300]{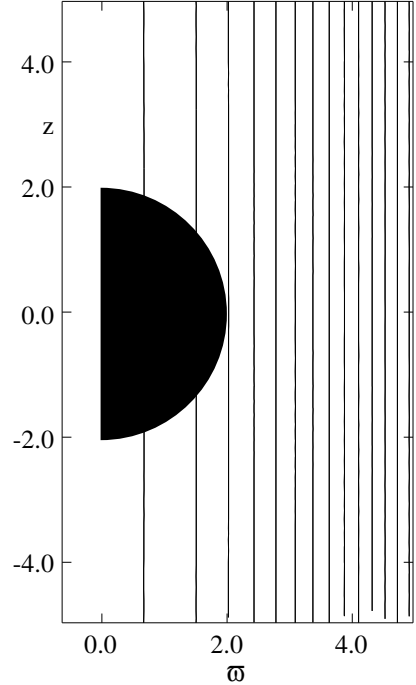}
\caption{Schwarzschild black hole in uniform magnetic field. The 
black circle shows the position of the event horizon and the continuous 
lines show the magnetic flux surfaces.}
\label{wald0}
\end{figure}

The typical values of $\check{B}$ and $u_b$ in vicinity of a supermassive black 
of $M \approx 10^8 M_\odot$ are $\check{B} \approx 10^4$Gauss and 
$u_b \approx 10^7 \mbox{erg}\,\mbox{s}^{-1}$. Hence, we have 
$$
  \nu_c \approx 0.1 W, \mbox{s}^{-1} \text{} 
  \nu_b \approx 10^{11} W^{-1} \mbox{s}^{-1}, 
$$
and 
\begin{equation} 
    \delta \approx 10^{12} W^{-2},   
\label{delta} 
\end{equation} 
where $W$ is the typical Lorentz factor of charged particles. 
Thus, the conductivity across the magnetic field is strongly suppressed
indeed, and the drift velocity is given by the familiar expression  

$$
\bv_d = \vpr{\cE}{\cB}/\check{B}^2. 
$$
For the Goldreich-Julian number density, 
$n_{GJ} \approx 0.1\, \mbox{cm}^{-3}$, we obtain    
$$ 
   \sigma_{\parallel} = 10^8 W^{-2} \mbox{s}^{-1} \text{} 
   \sigma_{\perp} = 10^{-16} W^2 \mbox{s}^{-1}, 
$$
whereas the macroscopic frequency is 

$$
 \nu_m = \frac{c^3}{2MG} \approx 10^{-3} \mbox{s}^{-1}. 
$$ 
Thus, one may confidently put $\cE_{\parallel}=0$ and ignore 
the cross-field conductivity current, 
$\sigma_\perp \cE_\perp =0$. 
This corresponds to the limit of force-free electrodynamics.     
Indeed, under these conditions 
$$
  \rho \cE +\vpr{j}{\cB}=  
  \rho \cE_\perp + \bj_d\!\times\!\cB = 0.   
$$ 
In this limit, the role of the $\cE_\perp$ component of electric field 
is reduced to driving of the drift current.

Inside a current sheet the cross-field conductivity can no longer be 
ignored. In fact, it has got to be governed by a self-regulatory mechanism
ensuring marginal screening of the electric field. Indeed, let us assume, 
for a moment, that the unscreened component of electric field is of the 
same order as the magnetic field.  
Then from equation (\ref{motion1}) it follows that 
 
\begin{equation} 
W \approx \sqrt{\frac{e\check{B}}{\sigma_T u_b}} \approx 10^6.   
\label{Gamma}
\end{equation} 
This implies $\delta \approx 1$ (see eq.\ref{delta} ) and, thus, similar  
conductivities both along and across the magnetic field. 
However, particles with such a high Lorentz factor are bound to launch 
copious pair cascades \cite{Bes,Hir-Oka}. This would dramatically 
increase the electric conductivity and lead to effective reduction of 
the unscreened component of the electric field. 
Thus, the electromagnetic field of a magnetospheric current sheet 
is expected to be very close to the state of marginal screening with 
\begin{equation} 
 \check{B}^2-\check{E}^2 = 0.
\label{screen3}
\end{equation}
(In fact, the electric field should remain slightly stronger than 
the magnetic field.) Since in this state, the typical  
particle's Lorentz factor is much smaller than the one 
given by eq.(\ref{Gamma}), one should have  
$$ 
\sigma_\parallel \gg \sigma_\perp, \quad
\check{E}_\perp \gg \check{E}_\parallel,  
$$ 
and
\begin{equation}
    \bj_d =  \rho \frac{\vpr{\cE}{\cB}}{\check{B}^2}.  
\label{jdrift2} 
\end{equation}

Summarizing, the cross-field conductivity is essentially zero 
in the main body of the black hole magnetosphere, where it is 
basically force-free, and dramatically increases in current sheets to 
ensure that the magnitude of $\cE_\perp$ exceeds the magnitude 
of the magnetic field only by a small margin.

\begin{figure}
\leavevmode
\epsffile[0 0 216 550]{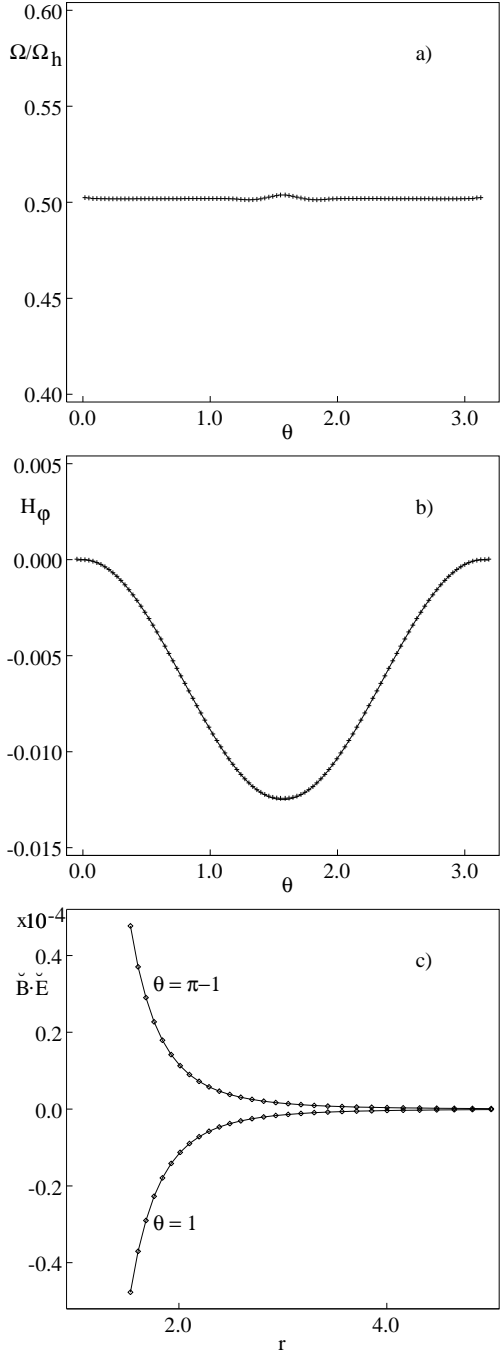}
\caption{Monopole field solution for $a=0.1$ and $B_0=1$ at 
time $t=50$. Panel a) shows 
the angular velocity of magnetic field lines at r=3. 
The perturbative solution of Blandford and Znajek gives 
$\Omega = 0.5 \Omega_h$, where $\Omega_h$ is the angular velocity 
of the black hole. Panel b) shows $H_\phi$ of the numerical solution 
(the crosses) and of the perturbative solution (the continuous line) at 
$r=3$. Panel c) shows $\spr{\cE}{\cB}$ along $\theta=1$ and
$\theta=\pi-1$. The small unscreened component $\cE_\parallel$ of 
electric field drives the conductivity current towards the black hole 
in the upper hemisphere and away from it in the lower hemisphere.} 
\label{mono}
\end{figure}

\section{Numerical study of black hole magnetospheres} 

To describe the black hole space-time we adopted the Kerr-Schild 
coordinates, $\{t,\phi,r,\theta\}$, which are free from the coordinate 
singularity at the event horizon hampering the Boyer-Lindquist 
coordinates (see Appendix B). This allows us to put the 
inner boundary of the computational domain inside the horizon and impose 
radiative boundary conditions on this boundary. 

All simulations described below are axisymmetric.
The computational grid is uniform in the $\theta$-direction with 
$\Delta\theta = \pi/n_\theta$, where $n_\theta$ is the total number of 
cells in this directions.  
The cell size in the r-direction, $\Delta r$, is determined via the condition 
of the same length in both directions, 
$\gamma_{rr}\Delta r^2 = \gamma_{\theta\theta}\Delta\theta^2$. 
The time step is determined by the Currant stability condition.

The details of the resistivity model used in these simulations 
are given in Appendix C3. 

\subsection{Wald's solution} 

Wald (1974) obtained the following vacuum solution for a rotating black 
hole immersed into a uniform magnetic field aligned with hole's 
rotational axis: 
\begin{equation}
   F_{\mu\nu}= B_0 (m_{[\mu,\nu]} +2a k_{[\mu,\nu]}), 
\label{wald} 
\end{equation} 
where $ k^\nu = \Pd{t}$ and $m^\nu = \Pd{\phi}$ are the Killing vectors 
of the Kerr spacetime. 

In the case of a Schwarzschild black hole eq.(\ref{wald}) reads  

\begin{equation}
    \bE=0, \qquad \bB=\frac{B_0}{2\sqrt{\gamma}} 
     \left(0,-\gamma_{\phi\phi,\theta}, \gamma_{\phi\phi,r}\right).   
\label{walds1}
\end{equation} 

Outside of the horizon this solution satisfies the plasma equations 
(\ref{divB}-\ref{Ampere},\ref{Ohm}) with zero $\rho$ and $\bj$. Indeed, 
from eq.(\ref{E3}) one finds 
$$
  \bD = -\frac{1}{\alpha} \vpr{\beta}{B}. 
$$  
Thus, $\bD_\parallel=0$ and no electric current is driven along 
the magnetic field. Provided $B^2-D^2 >0$ no electric current 
will be driven across the magnetic field lines either.  
From eq.(\ref{a12}) one has 
$$
   B^2-D^2 =B^2 \left( \frac{\alpha^2-\beta^2}{\alpha^2}\right) + 
                \left( \frac{\beta\gamma_{rr}B^r}{\alpha}\right)^2 
$$   
which is strictly positive outside of the event horizon. 
(Inside of the horizon $\alpha^2-\beta^2 < 0$ and $B^2-D^2$ 
becomes negative near the equatorial plane where $B^r=0$. However, 
these details do not effect the exterior solution.)   

The computational grid in this test problem has 100 cells in the
$\theta$-direction and 80 cells in the r-direction covering the domain
$r\in [1.4,29]$. The initial solution is described by eq.(\ref{walds1}). 
Figure \ref{wald0} shows the magnetic flux surfaces of 
the numerical solution at time $t=5$ as well as 
the flux surfaces of the exact steady-state solution (\ref{walds1}). 
In fact, these solutions are so close that one cannot see the difference
between them in this figure.

\begin{figure*}
\leavevmode
\epsffile[40 0 380 301]{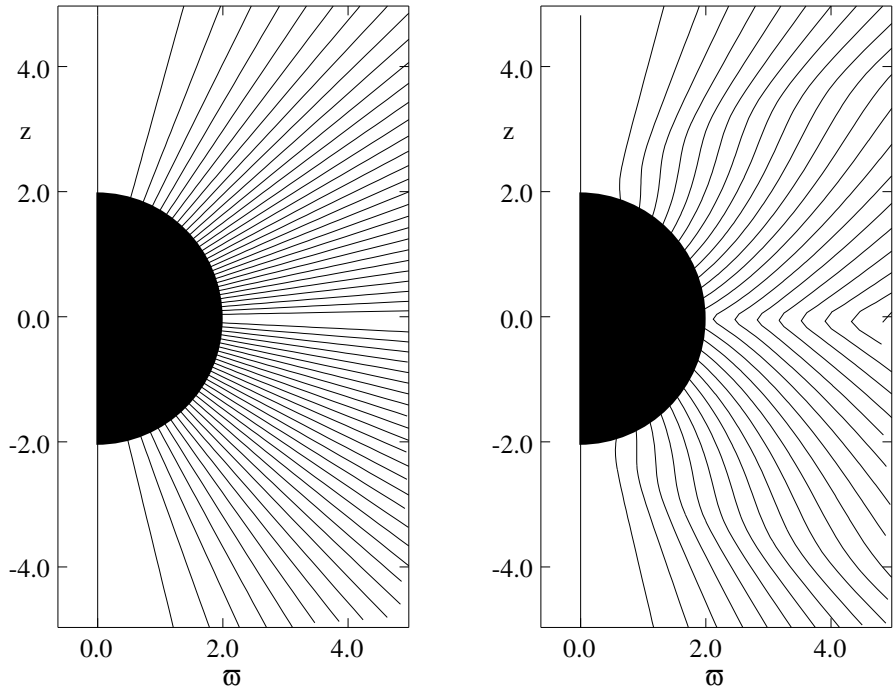}
\caption{Escape of the split-monopole magnetic field from a 
Schwarzschild black hole. {\it Left panel:} Magnetic 
flux surfaces of the split-monopole solution which was 
used as an initial solution in these numerical simulations. 
{\it Right panel:} Magnetic flux surfaces of the numerical 
solution at $t=5$. 
}
\label{smon}
\end{figure*}

\subsection{Blandford-Znajek's monopole solution.} 

Using a perturbation method Blandford and Znajek(1977) constructed a
perturbative analytic solution for a slowly rotating black hole with
monopole magnetic field that matches the flat space solution of
Michel(1973) at infinity and satisfies Znajek's ''boundary condition''
(see Sec. 6.2) imposed at the event horizon. The zero order solution 
is the one for a nonrotating black hole with purely radial magnetic 
field  

\begin{equation}
 B^r = B_0\sin\theta/\sqrt{\gamma}. 
\label{mon1}
\end{equation}
If $a \ll 1$ then the poloidal magnetic field is not expected to be very 
different from purely radial with $B^r$ given by (\ref{mon1}) and, thus, 
$B^r\sqrt{\gamma}$ remains constant along the field line. 
Moreover, the Znajek condition (eq.\ref{Zcon2}) reads  

\begin{equation}
 H_\phi = (\Omega-\Omega_h)\sin\theta B^r\sqrt{\gamma}, 
 \label{mon2}
\end{equation}   
where $\Omega_h=a/(r_+^2+a^2) \approx a/4$ is the angular velocity of the black 
hole. On the other hand, Michel's monopole solution in flat spacetime  gives 

\begin{equation}
 H_\phi = -\Omega\sin\theta B^r\sqrt{\gamma}. 
 \label{mon3}
\end{equation}   

Matching of $H_\phi$ given by (\ref{mon2}) with $H_\phi$ given by (\ref{mon2}) 
leads to 
\begin{equation} 
    \Omega=0.5\Omega_h=a/8 \text{and}  H_\phi =-\frac{a B_0}{8}\sin^2\theta 
\label{mon4}    
\end{equation}

In these simulations case we used $B_0=1$ and put $a=0.1$ because 
for this value of $a$ the corresponding force-free numerical 
solution is known to be very close to the perturbative one \cite{Kom01}. 
The computational grid in this test problem had 100 cells in the
$\theta$-direction and 150 cells in the r-direction covering the
domain $r\in [1.4,260]$.  We tried two very different initial
solutions for this problem. One of them describes the vacuum monopole 
field  (this solution was found numerically.) 
The other one has the same $\bB_p$ as in the
Blandford-Znajek solution and $\bE=0$. However, in both these cases 
the outcome was the same, the numerical solution gradually evolved 
towards the Blandford-Znajek one.  
Figure \ref{mono} shows the angular distribution of $\Omega$ and
$H_\phi$ at $r=3$  as  well as the distribution of the 
parallel component of the electric field in the northern and 
in the southern hemi-spheres at $t=50$.

\subsection{Split-monopole solution}      

Since the magnetic charges do not seem to exist, the monopole 
solution of Blandford and Znajek (1977) is rather artificial. 
A small modification,
namely alternation of the magnetic field direction in one of the
hemi-spheres, seems to overcome this problem. 
The result, known as the ``split-monopole'' solution, is still a
force-free solution but now with zero total magnetic flux. 
However, such a configuration is meaningful only if there is 
a massive perfectly conducting thin disc in the equatorial plane 
of the black hole. Otherwise, the equatorial current sheet is not 
sustainable -- the magnetic field is expected to reconnect and 
``fly away''. 
Indeed, this is what is observed in numerical simulations (see 
Figure \ref{smon}.)  In this problem $a=0.1$ and we use almost 
the same grid as in the Wald problem (Sec.5.1).

\subsection{The Magnetospheric Wald Problem for rotating black holes}

The Blandford-Znajek solution is the only global analytical solution
for magnetospheres of rotating black holes found so far and for this
reason it has been playing a key role in the development of the black hole
electrodynamics. One important property of this solution is that all
magnetic field lines penetrate the black hole horizon.  Macdonald
(1984) attempted to construct numerical steady-state solutions for
a more reasonable configuration of magnetic field where only a 
fraction of magnetic field
lines originates from the black hole itself. The remaining magnetic flux
splits between the field lines originating from the accretion disk and
the field lines passing through the gap between the hole and the
disc. In general, the angular velocity of magnetic field lines in
steady-state force-free magnetospheres has to be prescribed, so one
faces the task of setting physically sensible boundary conditions for
all these three different types of magnetic field lines.  In the case
of the field lines originating from the accretion disc the solution is
obvious. Their angular velocity is given by the angular velocity of
the disc at the foot points.  As for the other two kinds of magnetic
field lines, this task is less trivial.  In their solution, 
Macdonald and Thorne(1982) and later Macdonald(1984) appealed to the
existing analogy between the black hole horizon and a rotating
conducting sphere. They concluded that only the field lines
penetrating the event horizon rotate, whereas in the gap  
$\Omega=0$.

\begin{figure*}
\leavevmode
\epsffile[0 0 455 302]{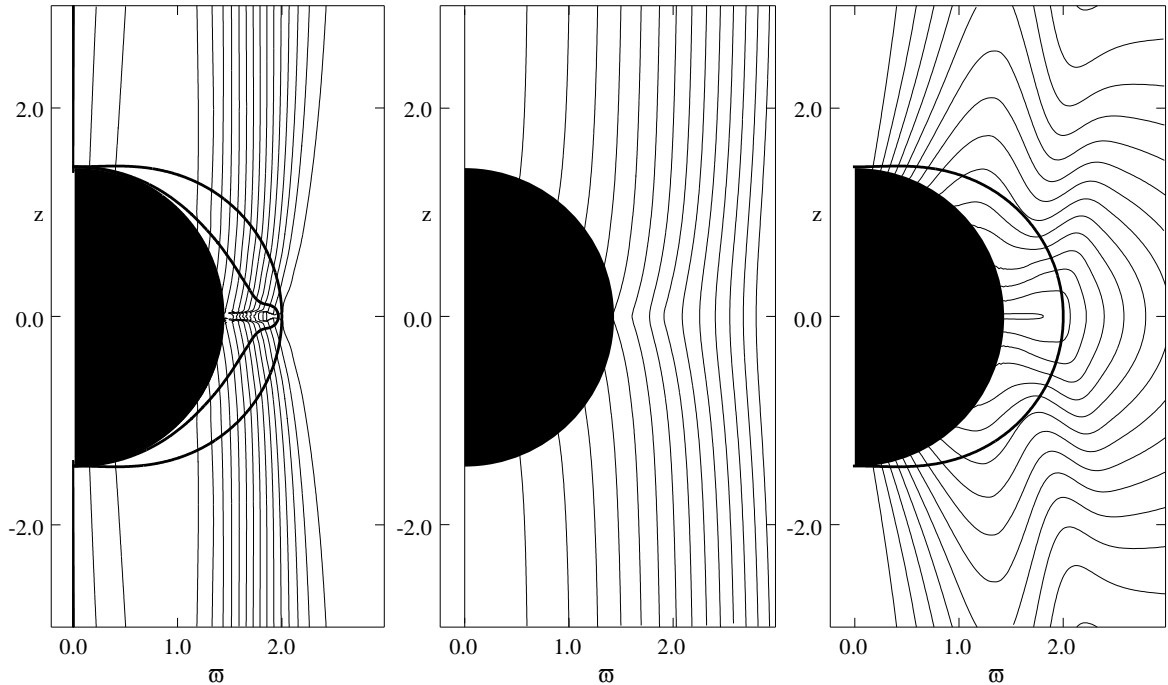}
\caption{Magnetospheric Wald problem. {\it Left panel:} 
The angular velocity of magnetic field lines. There are 15 
contours equally spaced between 0 and 0.67. The angular velocity first 
gradually increased towards the axis but then reaches a maximum and 
goes slightly down. The thick lines show the ergosphere (the outer line) 
and the inner light surface (the inner line). 
{\it Middle panel:} The magnetic flux surfaces. {\it Right panel:} 
The distribution of $(B^2-D^2)/\mbox{max}(B^2,D^2)$. 
There are 15 contour equally spaced 
between -0.12 and 1.0. This quantity monotonically decreases towards the 
current sheet in the equatorial plane within the ergosphere.  
The thick line shows the ergosphere.  
}
\label{wal-com}
\end{figure*}

A somewhat simpler problem is the magnetospheric (plasma-filled)
version of the Wald~(1974) problem for a rotating black hole 
(see also Sec.5.1). In this problem, just like in the problem 
considered by Macdonald~(1984), only a small fraction of magnetic
field lines penetrate the black hole horizon. If the analysis
of Macdonald \& Thorne~\shortcite{MT} was correct then only these 
field lines would be forced to rotate. 
Komissarov(2002b) tried to find a steady state force-free  
solution to this problem by means of time-dependent numerical 
simulations but failed. 
The numerical solution invariably evolved towards the state 
where $B^2-D^2$ turned negative inside the ergosphere. In fact, 
the solution seemed to indicate the development of 
a current sheet in the equatorial plane within the ergosphere with all
magnetic field lines penetrating the current sheet being forced 
to rotate in the same sense as the black hole.  
If this conclusion is correct, a critical revision of the current perception 
of the role of the event horizon in the black hole electrodynamics, as  
well as of the virtues of the Membrane Paradigm, is required. 
Thus, the magnetospheric Wald problem is an ultimate Rosetta Stone 
for the research into the black hole electrodynamics. 

To achieve high resolution within the ergosphere these simulations
were carried out using the multi-grid technique. We start with a
relatively low resolution grid and continue simulations until the
solution becomes more or less steady within $r=4$. Then the
resolution is increased by the factor of 2 and the simulations are
continued until new approximately steady-state solution is reached and
so on. During the grid refinement the numerical solution on the finer 
grid is found via interpolation.
The final grid has 800 cells in the $\theta$-direction
($\theta \in [0,\pi]$) and 1000 cells in the r-direction ($r \in
[0.9r_+, 110]$).  The initial solution is described by the same $\bB$
as in the vacuum solution of Wald (1974, eq.\ref{wald}) and has
$\bE=0$, which implies a non-rotating magnetosphere. 

\begin{figure*}
\leavevmode
\epsffile[0 0 432 200]{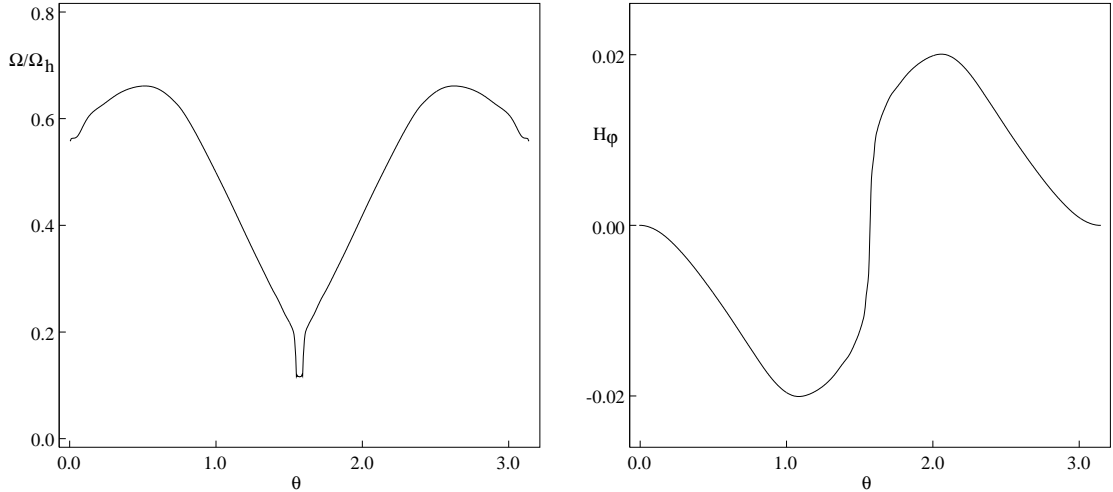} 
\caption{The distribution of the angular velocity of magnetic lines 
({\it left panel}) and $H_\phi$ ({\it right panel}) in the magnetospheric 
Wald problem at $r=1.8$.}
\label{wal-s}
\end{figure*}

Figure \ref{wal-com} shows the final solution, at $t=126$, for a Kerr
black hole with $a=0.9$. As suggested in Komissarov(2002b), a current
sheet is formed in the equatorial plane within the black hole
ergosphere.  This is clearly seen in the right panel of
fig.\ref{wal-com} which shows the distribution of
$(B^2-D^2)/\mbox{max}(B^2,D^2)$. Near the equator the predominantly radial electric
field is larger than the magnetic field and drives the electric
current across the poloidal magnetic field lines. Both the radial
component (the middle panel of fig.\ref{wal-com}) of magnetic field
and its azimuthal component exhibit a break in the equatorial plane on
the scale of the current sheet. The most important result is shown in
the left panel of fig.\ref{wal-com}: all magnetic field lines
penetrating the ergosphere are forced into rotation in the same sense
as the black hole irrespective of whether they eventually cross the
event horizon or not. Along these field lines there are outgoing
fluxes of both energy and angular momentum. Indeed, in steady
state force-free magnetospheres the angular momentum flux is
proportional to $H_\phi$ and the energy flux is proportional to
$\Omega H_\phi$ ( see eqs.\ref{a13},\ref{a14}.
Notice that $\Omega$ and $H_\phi$ are
the same in the Boyer-Lindquist as in the Kerr-Schild coordinates; 
see Appendix B). As 
one can see in fig.\ref{wal-s}, both these quantities are nonvanishing
along the field lines penetrating the ergosphere. Within the current
sheet the electromagnetic energy and angular momentum are not conserved
and, thus, the numerical results suggest that it is the current sheet
that supplies both the energy and the angular momentum for the
force-free magnetosphere above and below the sheet.

To show that this makes sense let us consider the sources of
energy and angular momentum in the symmetry plane, $\theta =\pi/2$. 
Because of the symmetry  $D_\parallel$ vanishes in this plane. Thus, we can ignore 
the dissipation due to $\sigma_\parallel$, and we can still introduce        
vector $\bOm$ to describe $\bD$ as in eq.(\ref{a5b}). In addition, 
the marginal screening of electric field implies $B^2=D^2$. 
The source terms in the Boyer-Lindquist coordinates have the same 
sign as in the Kerr-Schild coordinates and this allows us to 
utilize the Boyer-Lindquist coordinates to simplify calculations.    
 
The source term in the  energy equation (\ref{EC}) is 
$$ 
   {\cal W}_t = - \spr{E}{J}.
$$
Substituting the expressions
for $\bE$ and $\bJ$ from (\ref{E3},\ref{e5},\ref{Ohm}) into 
this equation one obtains

$$
    {\cal W}_t = - \alpha^2 \sigma_\perp D^2 - 
            \alpha\sigma_\perp \bD\!\cdot\!(\vpr{\beta}{B}). 
$$
Then, provided $\Omega < \Omega_z$, 
where $\Omega_z = -g_{\phi t}/g_{\phi\phi} $ 
is the angular velocity of the ``zero angular momentum observer'' 
(ZAMO, Bardeen et al.,1973), we have  

$$
    \bD\!\cdot\!(\vpr{\beta}{B})= -|\bbeta| B^2.   
$$ 
Hence, we obtain

\begin{equation} 
    {\cal W}_t = \alpha\sigma_\perp B^2(|\bbeta|-\alpha). 
\label{b6}  
\end{equation} 
Similar calculations lead to the expression 
for the angular momentum source, ${\cal W}_\phi$, in the 
equatorial plane 

\begin{equation} 
    {\cal W}_\phi = -(\vpr{J}{B})\!\cdot\! \bm = \alpha\sigma_\perp B^2.
\label{b7}  
\end{equation} 
Thus, provided the magnetosphere rotates slower then ZAMO, the resistive 
dissipation in the ergospheric current sheet can provide positive sources  
of electromagnetic energy and angular momentum.
Since, the inner light surface encloses the current sheet (see the left 
panel of figure \ref{wal-com}) 
the condition $\Omega<\Omega_z$ is satisfied in our numerical solutions.   
Obviously, this source of electromagnetic energy and 
momentum implies the existence of the same magnitude sink of energy 
and angular momentum for the radiation field. Thus, the energy transfer 
has to be essentially the same as proposed in \cite{Pun01}. Namely, the 
electromagnetic field pushes plasma particles into orbits with negative 
energy at infinity,  these particles push background photons into 
orbits with negative energy at infinity and finally these photons are  
swallowed by the black hole.        

As the grid resolution increases the current sheet begins to show sings of  
unsteady behavior. Whether this is only a numerical effect or an indication  
of the current sheets instability remains to be investigated.

\section{Discussion. The nature of Blandford-Znajek mechanism} 

The results of numerical simulations presented in Sec.5 allow us 
to draw the following two general conclusions concerning the electrodynamics of 
black hole magnetospheres. 

First of all, the rotational energy of black holes can 
indeed be extracted electromagnetically. In particular,   
the Blandford and Znajek solution is certainly stable and, hence, does not 
clash with causality. 

On the other hand, there is much more to the electrodynamics 
of black holes than it is proposed in the Membrane Paradigm and the analogy 
between a magnetized 
rotating conducting sphere and the black hole horizon is at least incomplete. In fact,
closer inspection of the causality arguments due to Punsly and Coroniti shows that 
it is the Membrane Paradigm which is most directly under attack. The Blandford-Znajek 
solution itself is involved mainly because it is widely considered as 
inseparably linked with the paradigm.

\subsection{The origin of torque}

\begin{figure*}
\leavevmode
\epsffile[0 0 440 240]{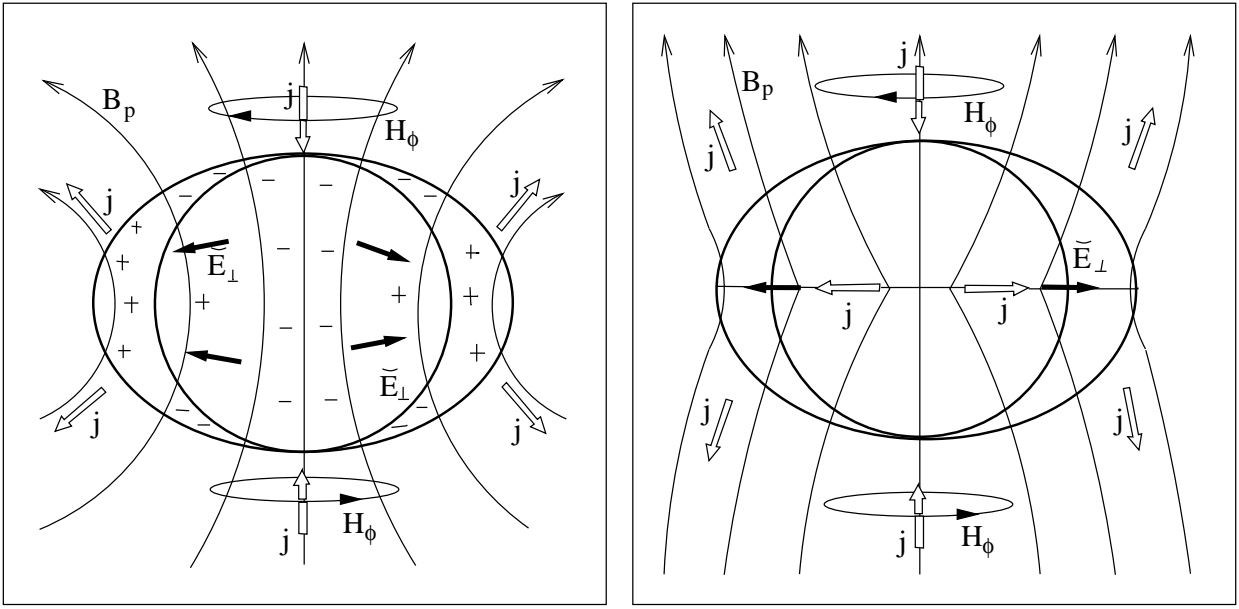}
\caption{{\it Left panel:} The initial currents driven by the gravitationally 
induced electric field within the ergosphere as it is being filled with 
pair plasma during the thought experiment described in Sec.6.1.1. 
{\it Right panel:} The electric current system of the steady-state Wald's 
magnetosphere.}
\label{inductor}
\end{figure*}

\subsubsection{Gravitationally induced electric field.}
It is well known that a black hole with zero total electric and magnetic charge 
cannot have its own magnetic field and, ultimately, any magnetic field penetrating the 
hole's ergosphere has to be supported by external currents. 
This fact alone makes 
black holes very different from magnetized stars like pulsars. 
Moreover, a steady state axisymmetric vacuum electromagnetic field 
cannot be used to extract energy and angular momentum of a 
rotating black hole. Indeed, the steady-state vacuum equations,  
$$
   \vcurl{H}=0, \quad \vcurl{E}=0, 
$$ 
ensure
\begin{equation}
H_\phi=E_\phi=0 
\label{o1}
\end{equation}
for axisymmetric configurations. Then from eqs.(\ref{S},\ref{L}) 
one finds that the poloidal components of the energy and the angular 
momentum flux vectors vanish as well. In order to extract energy 
and angular momentum the electromagnetic field has to be modified 
by magnetospheric charges and currents. However, in order to drive such 
currents and, perhaps, even to create charged particles via pair cascade in the first 
place \cite{BZ,Bes}, the electric field  should not be screened. 
That is at least one of the conditions (\ref{EB},\ref{a11})
has to broken in the vacuum solution. As it is well known, the vacuum 
solution due to Wald \shortcite{Wald} has this property but we need 
to know whether it is generic.   

{\bf Theorem.} {\it There are no steady-state axisymmetric 
vacuum solutions supported by remote sources of magnetic 
field that simultaneously satisfy both 
$$
    \spr{D}{B} =0 \text{and} B^2-D^2 >0  
$$  
along the magnetic field 
lines penetrating the ergosphere of a rotating black hole.} 

To prove this statement let us assume that electric 
currents of these 
sources can be arranged in such a way that  $\spr{D}{B}=0$ everywhere 
in the vacuum zone surrounding the black hole. 
Just like in the case of a force-free magnetosphere 
this condition ensures the existence of the vector 
field $\bOm=\Omega\bm$ such that $\Omega$ is constant the along magnetic field 
lines and 
$$
   \bE =-\vpr{\omega}{B}.
$$
The remote sources may be regarded as non-rotating relative to the
black hole and, thus, $\bE=0$ along the field lines penetrating the
ergosphere.  Then from eq.(\ref{E3}) one obtains

\begin{equation}
    \bD = -\frac{1}{\alpha}\vpr{\beta}{B}.  
\label{o2}
\end{equation}
Thus, even though there are no external sources of electric field, the
electric field is generated near the black hole as the result of the
inertial frame dragging effect manifested in the shift vector
$\bbeta$.  This is not entirely unexpected. Relative to any physical
observer located within the ergosphere, including FIDO, these external
sources do rotate.  In the Boyer-Lindquist coordinates $\bbeta$ is a purely 
azimuthal vector and $B_\phi=H_\phi/\alpha=0$ (see eq.109). This allows to find

\begin{equation}
    B^2-D^2 = \frac{\alpha^2-\beta^2}{\alpha^2}B^2.    
\label{o3}
\end{equation}

\noindent
Within the ergosphere $\alpha^2-\beta^2 <0$ and, hence, $B^2-D^2<0$. 
Thus, no matter how  strong is the magnetic field within the ergosphere,  
the gravitationally induced electric field is always stronger.  
This completes the proof.  

Obviously, these external sources may not be infinitely 
far away and strictly non-rotating. In such a case, $\Omega$ 
is not exactly zero and instead of eq.(\ref{o3}) one obtains 

\begin{equation}
   \alpha^2(B^2-D^2) = - B^2 f(\Omega,r,\theta),     
\label{o4}
\end{equation}
where
$$
 f(\Omega,r,\theta)=g_{\phi\phi}\Omega^2+2g_{t\phi}\Omega+g_{tt}   
$$ 
is the light surface function (see Appendix A). 
Thus, the condition $B^2-D^2>0$ is equivalent to the condition 
$\Omega > \Omega_{min}$ (see eq.\ref{t6a}). 
Since within the ergosphere $\Omega_{min}>0$ and reaches the value of 
$\Omega_h$ at the event horizon, this requires the remote 
sources to rotate really fast. In fact, unless the remote 
sources of magnetic field are located within the outer light surface
corresponding to $\Omega_{min}$, which may be theoretically possible 
but which is highly unlikely to occur naturally, this condition cannot 
be satisfied.

Thus, the vacuum field created within a black ergosphere by distant
sources must have unscreened electric field capable of driving
electric currents, provided the charged particles are injected there
somehow. The most popular mechanism involves an $e^-$-$e^+$ pair cascade
in strong electric and radiation field \cite{BZ,Bes}.  Let us consider
an initial vacuum solution of the kind discussed above, that is a
solution with $\spr{D}{B} = 0$ and $\Omega=0$, and figure out what 
occurs when plasma is injected (In fact, our initial numerical solutions 
in the Bladford-Znajek problem and the magnetospheric Wald problem 
are of this type.) 

First of all, FIDO's electric field $\cE=\bD$ given by (\ref{o2}) 
will drive the conductivity current across the magnetic field lines
within the ergosphere. This will result in the electric charge separation
and the drop of the electrostatic potential along the magnetic field lines
entering the ergosphere (see figure \ref{inductor}). The induced
parallel component of the electric field will drive the conductivity
current along the magnetic field lines leading to the dynamo of
$H_\phi$ according to the Ampere law $\vcurl{H} = \bJ$. 
As one can see in the
left panel of figure \ref{inductor}, the sign of the generated $H_\phi$ is
exactly the one which is required to slow down the black hole. However, we need
to verify that this is not a temporary phenomenon and a nonvanishing
$H_\phi$, and, hence, a nonvanishing poloidal current, will be a property of the
magnetosphere when it relaxes to a steady-state.

\subsubsection{Poloidal currents in static magnetospheres.} 
\label{prevsec}
Let us suppose that the magnetic field lines penetrating the ergosphere are
anchored to distant plasma of infinitely large inertia. Then the
created magnetosphere will eventually become static with $\Omega=0$,
$\bE=0$. From eq.(\ref{a12}) one finds that in a force-free region of
such magnetosphere

\begin{equation} 
B^2-D^2 = B^2 \frac{\alpha^2-\beta^2}{\alpha^2} + 
          \frac{1}{\alpha^2} (\spr{\beta}{B})^2 .   
\label{c1} 
\end{equation} 
In the Boyer-Lindquist coordinates this reads 

\begin{equation} 
\alpha^2(B^2-D^2) = B^2 (\alpha^2-\beta^2) + 
          \left(\frac{\beta^\phi}{\alpha}\right)^2 H_\phi^2.    
\label{c2} 
\end{equation} 
The difference between this result and equation (\ref{o3}) is entirely due 
to the azimuthal magnetic field created by the poloidal electric current.  
According to equation (\ref{a5b}), the growth of this component of magnetic 
field does not lead to the growth of $\bD$.

A number of important conclusions can be drawn from this simple result.
\begin{itemize} 
\item
First of all, inside the ergosphere $\alpha^2-\beta^2 <0$ and for
$B^2-D^2$ to be positive $H_\phi$ must be nonvanishing. 
This shows that screening of electric field by the injected plasma within 
the ergosphere implies generation of poloidal electric currents.  
These currents give rise to a nonvanishing $H_\phi$ and, thus, to a nonvanishing
flux of angular momentum.  Zero angular velocity of the magnetosphere
means zero Poynting flux from the hole (see eq.\ref{a14}). Thus,    
the total mass of the hole, $M$, remains constant but its irreducible 
mass, $M_{irr}$, grows due to the loss of angular momentum. One may
compare this case with a shorted out Faraday disk where its rotational
energy is being used only for Ohmic heating of the disc itself.

\item
Secondly, it makes no difference whether the magnetic field lines
penetrate the event horizon or not, {\it the only factor that matters is
whether they penetrate the ergosphere.}  The analogy between the event
horizon and the rotating conducting sphere emphasized in the Membrane
Paradigm does not reveal this important fact and, therefore, proves
to be of rather limited value.

\item
Finally, it explains the origin of the equatorial current sheet in the
magnetospheric Wald problem. Indeed, the magnetosphere is symmetric
relative to the equatorial plane, and for this reason alone $H_\phi$
must vanish in this plane.  Thus, the $B^2 - D^2 > 0$ condition cannot
be satisfied in the ergospheric part of the plane and the force-free
approximation breaks down. As we have seen in Sec.5.4 , this
current sheet serves as a localized source of angular momentum for
the surrounding force-free magnetosphere. Moreover, it provides the
required closure of the electric circuit at the black hole end 
(see the right panel of fig.\ref{inductor}). 
\end{itemize} 

\subsubsection{Poloidal currents in rotating magnetospheres} 
Since the  assumption of infinite inertia of the surrounding
plasma, made in Sec.\ref{prevsec}, is unrealistic and was made only 
for the sake of argument, the
outgoing flux of angular momentum will inevitably result in rotation
and, thus, extraction of the black hole energy as well.  The
equilibrium value of $\Omega$ depends of the details of the 
interaction.  In our simulations, the black hole was surrounded by
massless plasma and the equilibrium value of $\Omega$ was
determined by the rate of deposition of energy and angular momentum in
the surrounding space by means of propagating waves.

Even after reaching a force-free equilibrium the magnetospheric 
electric field does not become completely screened. 
Within the ergospheric current sheet, $\cE_\perp$ always
remains slightly stronger than the magnetic field and keeps driving
the cross field conductivity currents, thus, sustaining the
potential drop along the magnetic field lines entering the current
sheet.  In the force-free region itself 
it is the small residual component of $\cE_\parallel$ that drives the poloidal 
conductivity currents (see fig.\ref{mono}c), just as it occurs in the wires 
connected to the Faraday disc.

It is also quite clear why the magnetic field lines remaining outside 
of the ergosphere along their entire length do not rotate. 
Although the vacuum Wald solution has an unscreened  $\cE_\parallel$ component of 
the electric field which can trigger pair production and  
drive poloidal currents for a while, the injected charges eventually 
redistribute along the field lines and screen $\cE_\parallel$. 
Then, because $\cE_\perp$ is too weak (see eq.\ref{c2}), even in the 
equatorial plane, to drive the cross field current,  
the poloidal currents  die out completely. 
The force free conditions are met even when $H_\phi=0$.

The spatial grids of both the Boyer-Lindquist and the Kerr-Schild 
coordinates, though stationary relative to distant inertial observers 
at rest relative to the black hole, rotate superluminally 
relative to any local physical observer located within the ergosphere.  
This property of black hole space-time is crucial in determining the 
properties of their magnetospheres. To illustrate the point 
let us consider a steady-state magnetosphere rotating with angular 
velocity $\Omega$ in flat space-time (we assume that some particle injection 
mechanism ensures plentiful supply of electrical charges required to screen 
the electric field.) Let us introduce a spherical   
grid rotating with the same angular velocity. The corresponding coordinate
transformation   
 
$$ 
t=t',\ r=r', \phi = \phi '-\Omega t',\ \theta=\theta ',     
$$
where the primed coordinates refer to the nonrotating grid,  
results in the metric form 

\begin{equation}  
  ds^2=(\beta^2-\alpha^2)dt^2 +2\beta_\phi d\phi dt + dr^2+
        r^2( \sin^2\!\theta d\phi^2 + d\theta^2)
\label{c3} 
\end{equation} 
where 
$$
  \alpha=1,\ \beta^2=\Omega^2\sin^2\!\theta r^2,\ 
  \beta_\phi=\Omega\sin^2\!\theta r^2. 
$$ 
We notice that this metric is rather similar to the Kerr metric 
as the surface $r\sin\theta = \varpi_{c}=1/\Omega$ separates the inner 
region where the time coordinate is time-like from the outer region where 
it is space-like (obviously, $\varpi_{c}$ is the cylindrical radius of the 
light cylinder.) Relative to the rotating grid the magnetosphere is 
static and, thus, $\bE=0$. Calculations similar to those leading to 
eq.(\ref{c2}) give us  

\begin{equation} 
B^2-D^2 = B^2 (1-(\varpi/\varpi_{c})^2) + 
          \Omega^2 H_\phi^2.    
\label{c4} 
\end{equation} 
Thus, high conductivity ensures $H^{'}_\phi=H_\phi\not=0$ outside of 
the light cylinder. The similarity with black hole magnetospheres is 
startling.  

The result (\ref{c4}) serves another goal. It shows  
that $H_\phi$ must remain nonvanishing in rotating magnetospheres 
of black holes for any value of $\Omega \not=0$.   

\subsubsection{ Where is the unipolar inductor of a Kerr black hole?} 
    
In the case of the Faraday disc, which is a classical example of a unipolar 
inductor, electrons are forced to participate in the disc rotation via 
collisions with other disc particles. This provides the electromotive 
force, $q\vpr{v}{B}$, which results in electric charge separation and, 
hence, the voltage drop between the disc rim and its centre. 
When the disc is used as battery in a closed electric circuit the electromotive 
force continues to push electrons against the electric force and across
the magnetic field lines. This is essential for sustaining the 
potential drop and for providing the current closure.  
Spinning magnetized cosmic objects like stars or accretion discs 
generate an electric field in the very much same way.  
Such a field is often described as {\it rotationally induced}.   

Although, the Membrane paradigm invites to treat black holes in a similar 
fashion, in reality they are rather different. Both in the Blandford-Znajek 
solution and in our solution to the magnetospheric Wald problem there is no any 
massive conducting rotating object and, thus, no usual electromotive force 
driving electric current over its surface. In spite of this curious 
feature the potential drop across the magnetic field lines still exists 
and the electric currents still flow. 

The answer to this paradox seems to reside in another peculiar property 
of rotating black holes - contrary to our everyday experience the electric 
charge separation in not the only way of creating stationary electric field 
in their vicinity. The vacuum solution found by Wald(1974) shows that such 
an electric field can be induced {\it gravitationally}.  
Thus, the usual electromotive force 
is not required at least for initial charging of the black hole ``battery''.   
Punsly and Coronity (1990a) argued that such electric field can 
not sustain stationary electric currents. Their first point       
is based on the change of sign of the parallel component of the electric 
field in the exact Wald solution. Indeed, this seems to be inconsistent with 
the electric current flowing along the poloidal magnetic field line in one 
particular direction.  
However, the electric field modified by the magnetospheric charges does not have 
to retain this property. In their second point, they argue that the mere fact 
of existence of any initial electric field does not matter very much. For example, 
a capacitor can drive only a transient electric current. 
Unless a battery is continuously recharged via the action of some electromotive force 
it cannot drive a stationary electric 
current. In other words, the initial transient currents can redistribute 
electric charge in the black hole magnetosphere in such a way 
that the electric field becomes completely screened and can no longer drive 
the electric currents. However, as we have shown above, such a final state is not
possible within the black hole ergosphere. 
Black holes do not allow stationary solutions with screened electric field and 
vanishing poloidal electric currents.   
  
Thus, the electrodynamics of rotating black holes is very different from 
electrodynamics of usual magnetic rotators and their batteries operate 
on other principles than the classical unipolar inductor. 
The key role played by the ergosphere in black hole 
electrodynamics allows us to call it the ``driving force'' of the 
Blandford-Znajek mechanism.

\subsection{Causality arguments} 

Apart from the absence of the proper unipolar inductor 
Punsly and Coronity \cite{PC,Pun01} criticized the 
electromagnetic mechanism on the ground of causality principle.   
The key point of their criticism is concerned with the role of Znajek's  
horizon condition in the Blandford-Znajek solution (Blandford \& Znajek
1977, Znajek 1977; see also Sec.5.2.) Since Blandford and Znajek
employed the Boyer-Lindquist coordinate system, the horizon appears as 
a singular boundary of their spatial domain and the Znajek condition 
appears as a boundary condition.  In fact, this perception of the  
Znajek's condition is fully accomodated in the Membrane paradigm, where this
condition is used to endow the event horizon with the properties of rotating 
conducting surfaces, reintroducing the missing unipolar inductor in 
a somewhat ghostly form.  

Punsly and Coroniti \shortcite{PC} convincingly argued that this analogy 
must be wrong. They pointed out that the BZ-solution describes not just an  
outgoing wind, as in the case of pulsar winds, but also an ingoing wind 
that passes though its own critical surfaces
(see also Takahashi et al. 1990 and Appendix A.) 
As the result, the event horizon is causally disconnected from
the outgoing wind and, thus, not suitable for imposing boundary 
conditions that would determine the properties of the outgoing wind. 
This made Punsly and Coroniti \shortcite{PC} to conclude that
although the Blandford-Znajek solution may be a valid solution of
steady-state equations it must be unstable and thus the  
time-dependent solution would never evolve towards it.  
However, the results of numerical simulation presented in \cite{Kom01} 
and in this paper show just the opposite -- the Blandford-Znajek solution 
is, in fact, asymptotically stable and, hence, causal.
 
The answer to this controversy seems to be very simple. 
Znajek's horizon condition is not a boundary condition after all.
Indeed, Znajek derived the horizon condition from a very natural requirement. 
Namely, the electromagnetic field at the event horizon had to be 
non-singular when measured by a local free-falling observer. 
However, in the limit of force-free electrodynamics, the
event horizon coincides with the fast critical surface of the ingoing
wind.  This immediately follows from the fact that the fast wave of
force-free electrodynamics propagates with the speed of light
\cite{Kom02a}. All this suggests that Znajek's condition is in fact a 
regularity condition at the fast critical point of the ingoing wind,  
obtained by Znajek in a somewhat unconventional way. 

There is an important 
difference between a boundary condition and a regularity condition. 
Boundary conditions are set on the boundaries of computational 
domain and select particular solutions to both steady-state and 
time-dependent problems. 
Regularity conditions apply only to steady-state problems 
when one is looking for solutions passing smoothly through critical 
points, at which the steady-state equations change their type.

In the Kerr-Schild coordinates, where there in no coordinate singulatity at the 
horizon, the critical nature of the Znajek condition becomes very clear. 
Let us consider a steady state force-free solution smoothly passing through
the event horizon. In such a solution the angular velocity of
magnetic field, $\Omega$, and $H_\phi$ are constant along the magnetic
field lines. Following Weber \& Davis(1967) we can use these constants
in order to find the relationship between $B^\phi$ and $B^r$ along a given
magnetic field line. First, we eliminate $\bD$
from eq.(\ref{H3}) using eq.(\ref{a5b}) to obtain 
$$
   \bH = \alpha \bB +\frac{1}{\alpha} \left[
    \bmath{\omega}(\spr{\beta}{B})- \bB (\spr{\beta}{\omega} +
    \beta^2) + \bmath{\beta}(\spr{\beta}{B}) \right] .
$$
From this equation we derive, after rather involved calculations, the
following result:

\begin{equation} 
   B^\phi = \frac{\alpha H_\phi - B^r\sin^2\theta (2r\Omega-a)}
                 {\Delta \sin^2\theta}.
\label{Zcon1} 
\end{equation} 
The denominator of the right hand side of this equation vanishes at the
event horizon, where $\Delta=0$.  For $B^\phi$ to remain finite the
numerator has to vanish as well and we obtain

\begin{equation} 
   H_\phi = \frac{\sin^2\theta}{\alpha_+}(2r_+ \Omega-a) B^r =
   \frac{(2r_+ \Omega-a)\sin\theta}{r_+^2+a^2\cos^2\theta}
   A_{\phi,\theta} ,
\label{Zcon2} 
\end{equation} 
which is exactly the Znajek condition (Notice that $H_\phi$, $A_\phi$,
and $\Omega$ are the same in the Kerr-Schild coordinates as in the
Boyer-Lindquist coordinates; see Appendix B) 
Thus, Znajek's ``boundary condition'' is
indeed a regularity condition at the fast critical point of the
ingoing wind. In MHD approximation, in the case where the particle 
inertia is not negligibly small, the fast surface is located outside 
of the event horizon and Znajek's horizon condition is no longer needed 
to determine the steady-state wind solution \cite{Bes-Kus}.  

In time, the dispute over causality of the Blandford-Znajek solution, 
and causality of the electromagnetic mechanism as such, 
gradually evolved into a dispute over the role played by different waves 
in the development of the system of poloidal currents in black hole 
magnetospheres, e.g. \cite{Bland91,Pun96,Pun01,Bland02}. 
The origin of this slight change of subject resides in the fact
that in the Membrane paradigm the inner boundary conditions are
imposed not at the actual event horizon but at the so-called
``stretched horizon'', located just outside of the actual one
\cite{TPM}. This allows communication between the
stretched horizon and the outgoing wind by means of fast waves
giving hope of salvaging the perception of the event horizon as a unipolar 
inductor of black holes  \cite{Bland91,Bland02}. 
Punsly \shortcite{Pun96,Pun01} dismissed this idea showing that in 
order to provide the necessary adjustment of poloidal currents the 
stretched event horizon would have to communicate with the outgoing wind 
by means of Alfv\'en waves. However, this is not possible as the inner 
critical Alfv\'en surface in the Blandford-Znajek solution is located well
outside of the stretched horizon. 

The conclusion of Punsly \shortcite{Pun01} on the crucial role of 
Alfv\'en waves in establishing the global current system seems to be  
correct in general.  This is particularly easy to see in the limit of 
force-free electrodynamics. In this approximation the electric charge and 
the electric current density can only be changed by hyperbolic waves. 
The fast waves cannot do this as they have identical properties to 
linearly polarised waves of vacuum electrodynamics \cite{Kom02a}. The only
other option is the Alfv\'en waves. 

However, once we have seen that 1) the event 
horizon does not play the role of a unipolar inductor, 2) the Znajek's 
horizon condition is just the usual regularity condition, and 3) the key 
role in the electrodynamic mechanism is played by the black hole ergosphere,  
this dispute has to be somewhat redirected. What we need to verify 
is that the ergosphere is causally connected with the outgoing wind. 
In fact, as it is shown in Appendix A1, the inner Alfv\'en surface is 
always located inside the ergosphere, and our numerical simulations are 
fully consistent with this result (see fig.\ref{wal-com},
\ref{lsurf}). All these allow us to conclude that there is no causality 
clash associated the electrodynamic mechanism in general and        
with the Blandford-Znajek solution in particular.

\section{Conclusions}

\begin{itemize} 

\item The 3+1 equations of black hole electrodynamics can be written 
in more general and simpler form than the classical formulation of
Thorne and Macdonald \cite{TM,MT,TPM}. 
Our equations are very similar to the classical equations of electrodynamics 
in matter and hold in any system of coordinates which delivers time-independent 
metric form, e.g. Kerr-Schild coordinates.

\item  
The inertia of magnetospheric plasma is not an essential element in the mechanisms 
of extraction of rotational energy of black holes involving electromagnetic field. 
We have seen no indications of any major deficiency of the electrodynamic model 
which would suggest a need for the more complex MHD model,
at least at the black hole end. 
The force-free approximation, however, may break down locally via formation of 
dissipative current sheets. We have constructed a simple model for radiative 
resistivity based on the inverse Compton scattering of background photons and 
used this model to show the development of the equatorial current sheet within 
the ergosphere of a black hole with a dipolar magnetic field. This current   
sheet supplies energy and angular momentum for the surrounding force-free 
magnetosphere. 

Having said that, we need to stress that the role of particle inertial has not 
been considered in this paper and requires further investigation. 
For example, one would expect the inertial effects to become important in the 
remote part of the outgoing wind, like they seem to do in the winds from 
neutron stars, e.g. \cite{Mestel,KenCor}.  
       
\item  The gravitationally induced electric field is the ultimate cause of 
the poloidal currents in the black hole magnetospheres. 
Within the ergosphere this field cannot be 
screened by any static distribution of electric charge.  
This makes a ``magnetized'' black 
hole very different from the classical unipolar inductor (or any other battery),   
where the potential difference between terminals is created and sustained by 
the electromotive force which first separates electric charges and then drives 
electric currents against the electric field. 
In fact, within the ergospheric current sheet the 
cross-field current flows along the electric field but not against it. 
The special role played by the ergosphere allows us  
to call it  the ``driving force'' of the Blandford-Znajek mechanism. 
Since the same applies to the otherwise rather different Penrose mechanism 
\cite{Penrose}, this suggests that it is the existence of the ergosphere that 
makes possible energy extraction from a black hole in any form.            

\item The Blandford-Znajek solution in particular, and the electromagnetic 
mechanism of extraction of rotational energy of black holes in general, 
do not clash with causality. First of all, 
the Znajek horizon condition is not a boundary condition but a regularity condition 
at the fast critical point. This perfectly justifies its utilization in the 
Blandford-Znajek monopole solution. Secondly, the ergosphere is causally connected 
to the outgoing wind by means of both fast and Alfv\'en waves. This conclusion 
is strongly supported by numerical simulations that show that the Blandford-Znajek
monopole solution is asymptotically stable. 

\item Our results fully agree with the conclusion of Punsly and Coroniti 
\cite{PC,PC1} and Beskin \& Kuznetsova \shortcite{Bes-Kus} on the superficial 
nature of the interpretation of the event horizon as a unipolar inductor 
of black holes. The failure of the Membrane paradigm to 
predict the outflow of energy and angular momentum along all magnetic field lines 
penetrating the ergosphere has clearly exposed its limitations.  
Although the analogy with conducting sphere is based on mathematically sound 
grounds, it does not account for all important properties of black hole 
electrodynamics and does not reveal the true physical nature of the Blandford-Znajek 
mechanism.   

\end{itemize}

\section{Acknowledgments}
I thank Brian Punsly for the highly motivating discussion of the 
Blandford-Znajek mechanism over the Internet and Roger Blandford for 
his useful comments on this subject. I also acknowledge the limited 
support of this research from PPARC via  ``A rolling programme of 
theoretical astrophysics research in Leeds''. Finally, I would like 
to thank the anonimous referee for useful suggestions concerning 
the notation in the 3+1 electrodynamics.

\appendix 

\section{Light surfaces}

The physical significance of the light surfaces in the black hole magnetospheres 
is well understood and indisputable. However, the author of this paper 
could not find any more or less complete and systematic description of their 
properties. This section is intended to fill the gap and to provide a convenient 
reference for the reader of the paper.

\subsection{Basic properties of light surfaces}

Consider a point orbiting a Kerr black hole with angular velocity
$\Omega$. The type of its world line depends on the sign of the function

\begin{equation}
 f(\Omega,r,\theta)=g_{\phi\phi}\Omega^2+2g_{t\phi}\Omega+g_{tt} .  
\label{t0}
\end{equation}
It will be time-like (subluminal rotation) if $f<0$,
space-like (superluminal rotation) if $f>0$, and null if $f=0$.
Similarly, a magnetosphere, rigidly rotating with angular
velocity $\Omega$,  is divided into the regions of subluminal
and superluminal rotation depending on the sign of $f(\Omega,r,\theta)$.
The separating surfaces, defined via

\begin{equation}
  f(\Omega,r,\theta)=0,  
\label{t1}
\end{equation}
are called ``the light surfaces''.
Here we derive the basic properties of light surfaces of magnetospheres
with $0 < \Omega <\Omega_h$, where $\Omega_h = a/2r_+$ is the
angular velocity of the black hole (without any loss of generality
we assume that $a>0$.)  

\begin{enumerate}

\item {\it The light surfaces are symmetric relative to the
equatorial plane.}

This property immediately follows from the
symmetries of the metric tensor.
\vskip 0.2cm  

\item {\it There exist only two light surfaces, the inner one and 
the outer one.
Inside of the inner surface and outside of the outer surface
the rotation of the magnetosphere is superluminal.}

Consider the case $\Omega=0$. Then 
\begin{equation}
  f(\Omega,r,\theta)=g_{tt}
\label{t1a}
\end{equation}
and, thus, the only light surface is the ergosphere of the black hole.
Because $g_{tt}$ is positive inside of the ergosphere and negative outside 
of it, the rotation inside of this light surface is superluminal. Since
$f(\Omega,r,\theta)$ is analytical, this light surface will continue
to exist for $\Omega>0$, though its location may differ. We shell call
it the ``inner light surface''.  Since along the rotational axis
eq.(\ref{t1}) is reduced to $g_{tt}=0$ for any $\Omega$, this surface always
includes the point $(r,\theta)=(r_+,0)$.

If $0 < \Omega \ll 1$ then another light surface emerges from infinity.
Indeed, for $r\gg1$ 
\begin{equation}
   f(\Omega,r,\theta)= r^2\sin^2\theta\Omega^2 -1 
\label{t2}
\end{equation}   
and, thus, eq.(\ref{t2}) has the solution
\begin{equation}
  r=\frac{1}{\Omega\sin\theta}, 
\label{t3}
\end{equation}
which moves to infinity as $\Omega \to 0$. We shell call this light
surface the ``outer light surface''. From (\ref{t2}) it follows that outside 
of this surface the magnetosphere rotates superluminally. 
Notice, that according to  eq.(\ref{t2}) no other light surface
can continue to infinity for any value of $\Omega$.

Since $f(\Omega,r,\theta)$ is analytical, an additional light surface
can only 1) bifurcate from a point where $f(\Omega,r,\theta)$ has
a local minimum or maximum,  or 2) appear as a result of splitting of
already existing light surfaces. To exclude the first option we
notice that
\begin{displaymath} 
  \begin{array}{rl}
   \rho^4 \partial f/\partial \theta =  & 
       [(\Delta\rho^4+2r(a^2+r^2)^2)\Omega^2 - \\
       & -4ar(a^2+r^2)\Omega +2a^2r
       ]\sin2\theta , 
  \end{array}
\end{displaymath} 
which is positive for $r>r_+$ with exceptions of the rotational
axis and the equatorial plane where it vanishes. However, the only
point of the rotational axis where $f(\Omega,r,\theta)$ ever vanishes
belongs to the inner light surface. Similarly, there are only
two locations on the equatorial plane where $f$ ever vanishes and
those belong to the inner and the outer light surfaces for all
values of $\Omega$. Indeed, along the equator
$f=s(\Omega,r)/r$ where

\begin{equation}
   s(\Omega,r) = (r^3+a^2r+2a^2)\Omega^2-4a\Omega +2-r.    
\label{t4}
\end{equation}
Since $s$ is a cubic polynomial in r, it has at most three roots.
The extremes of the cubic are given by
\begin{equation}
  r^2=\frac{1}{3} \left( \frac{1}{\Omega^2} -a^2 \right), 
\label{t5}
\end{equation}
which shows that one of the roots is always negative. For
$\Omega\ll 1$ the two other roots are $r_{in}=2$ and
$r_{out}=1/\Omega$.
Obviously they belong to the inner and the outer light surfaces.
This also proves that no additional light surfaces can appear via
splitting of the inner or the outer ones.

Finally, on the event horizon 

\begin{equation}
   f(\Omega,r,\theta) = \frac{4r_+}{\rho_+^2} 
              \sin^2\theta(\Omega-\Omega_h)^2 , 
\label{t6}
\end{equation}
showing that no light surface can bifurcate from the event horizon
either.
\vskip 0.2cm  

\item{\it Inside the inner light surface the rotation is ``too slow'' 
whereas outside of the outer light surface it is ``too fast''.} 

The condition of subluminal rotation, $f(\Omega,r,\theta)<0$, 
leads to the following constraint on $\Omega$: 

\begin{equation}
     \Omega_{min} < \Omega < \Omega_{max},  
\label{t6a}
\end{equation}
where 
$$
  \Omega_{min} = \Omega_z-\sqrt{\Omega_z-g_{tt}/g_{\phi\phi}},  
$$
$$
  \Omega_{max}=\Omega_z+\sqrt{\Omega_z-g_{tt}/g_{\phi\phi}},  
$$
where $\Omega_z=-g_{t\phi}/g_{\phi\phi}$ is the angular velocity
of the local zero angular momentum observer (ZAMO).
On the horizon $\Omega < \Omega_h=\Omega_{min}$ and by continuity 
$\Omega < \Omega_{min}$ in the whole of the inner superluminal region. 
Similarly, for $r\to\infty$ one has 
$\Omega_{max} < 1/r\sin\theta < \Omega$ and by continuity 
$\Omega > \Omega_{max}$ in the whole of the outer superluminal 
region.   
\vskip 0.2cm  

\item {\it The inner and the outer light surfaces do never have common
points.}

Since, as we have established above, $\partial f/\partial\theta>0$ in the
space between the rotational axis and the equatorial plane, all we
need to show is that the root $r_{in}$ of eq.(\ref{t4})  
is always less than the root $r_{out}$.
We already know that this is true for $\Omega \ll 1$ and, hence, it
is sufficient to show that these roots of eq.(\ref{t4}) never merge.
From eqs.(\ref{t4},\ref{t5}) one finds the merger condition
$$
    \xi(a,\Omega) =(1+a\Omega)^3+27a\Omega^3-27\Omega^2=0. 
$$
It is easy to see that $\partial\xi/\partial a > 0$. Moreover, direct
calculations show that $\xi(a,\Omega_h(a)) \ge 0$, where the 
equality holds only if $a=1$. These ensure $\xi(a,\Omega)>0$
for $0<\Omega<\Omega_h(a)$. 
\vskip 0.2cm  

\item {\it The inner light surface is located between the ergosphere
and the event horizon.}

At the ergosphere
$$
    f(\Omega,r,\theta)=g_{\phi\phi}\Omega(\Omega-2\Omega_z). 
$$
For $0<\Omega\ll 1$ one has $\Omega < \Omega_z$ and $f<0$. Thus,
the inner light surface shifts inside the ergosphere as $\Omega$
turns positive. A light surface can also make contact
with the ergosphere when $\Omega=2\Omega_z$ but that can only be
the outer light surface. Indeed, simple calculations show that
$\partial\Omega_z/\partial r<0$ for $r>r_+$. Thus, along the radial ray
passing though the point of contact $f(\Omega,r,\theta)$ is positive
all the way from the ergosphere to infinity.

The conclusion that the inner light surface always remains 
outside of the event horizon simply follows from eq.(\ref{t6}). 
\vskip 0.2cm  

\item {\it The coordinate $r$ of the inner light surface
monotonically increases with $\theta$ for $0<\Theta<\pi/2$. 
For the outer light surface there holds exactly the opposite.}

Direct inspection of the lights surfaces found in the limit $\Omega\ll 1$ 
(see eq.\ref{t1a} and eq.\ref{t3})
immediately shows that they have this property. Thus, 
one need only to verify that $dr/d\theta $ does not vanish anywhere on 
the critical surfaces with the exceptions of the equatorial plane 
and the symmetry axis.  Since
$$
    \oder{r}{\theta} = -\pder{f}{\theta} /
                        \pder{f}{r}
$$
this follows from the fact that $\partial f/\partial\theta$ does not vanish
for $0 < \Theta < \pi/2$ (see the proof of property (ii)). 

\end{enumerate}
The geometrical properties of light surfaces are summarized in 
figure \ref{lsurf}. 

\begin{figure}
\leavevmode
\epsffile[0 0 216 216]{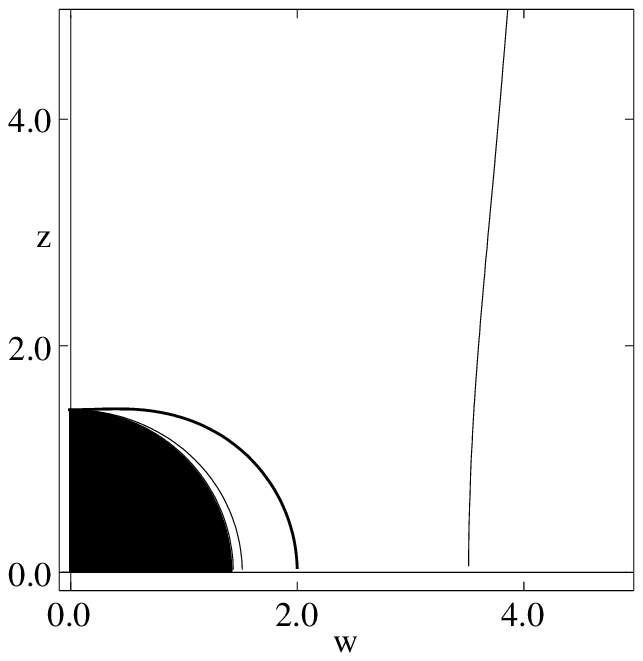}
\caption{The light surfaces (thin lines) of a rigidly rotating magnetosphere.    
Here $\Omega =0.7\Omega_h$ and $a=0.9$. 
The thick line shows the ergosphere.} 
\label{lsurf}
\end{figure}

\subsection{Light surfaces and magnetically driven winds}  
Consider a force-free region of a steady-state magnetosphere. 
As we have established above, each magnetic field line that penetrates 
the black hole ergosphere is forced to rotate with angular velocity 
$0 < \Omega < \Omega_h$ and may cross the light surfaces corresponding 
to this value of $\Omega$. Let us show that in the inner superluminal region 
charged particles are forced to move towards the black hole whereas in the 
outer superluminal region they are forced to move away from it. 
To simplify calculations we shell use the Boyer-Lindquist coordinates and 
assume, without any loss of generality, that the magnetic field line 
under consideration is outgoing ($B^r>0$).

Any such particle slides freely along the magnetic field line  
and participates in the drift motion across the field line. 
Thus, its velocity vector is 

\begin{equation}   
  \bv = \vpr{D}{B}/B^2 +\kappa\bB,
\label{f1} 
\end{equation} 
(for simplicity sake we ignore particle's gyration.) The radial component of $\bv$
is limited from above and from below via the condition $v^2<1$. 
In the limit $v^2=1$ one has $\bv=\bv_\pm$, where     

\begin{equation}   
  \bv_\pm = \frac{1}{B^2}(\vpr{D}{B}\pm\bB\sqrt{B^2-D^2}). 
\label{f2} 
\end{equation} 
Using eqs.(\ref{E3},\ref{a5}) to express $B^2-D^2$ and $\vpr{D}{B}$ 
in terms of $\bB$ and $\Omega$ one obtains the following result for 
the poloidal component of $\bv$: 

\begin{equation}  
\bv_{p_\pm}=\eta_\pm \bB_p,  
\label{f3} 
\end{equation} 
where
$$
  \eta_\pm = \frac{1}{B^2} \left(-l  
          \pm\sqrt{ l^2 - \frac{B^2}{\alpha^2} f(\Omega,r,\theta)} 
     \right),
$$
$$ 
   l= \frac{(\Omega-\Omega_z)H_\phi}{\alpha^2}. 
$$ 
Given the assumed  direction of $\bB_p$, $\bv_{p_+}$ has the 
highest possible and $\bv_{p_-}$ has the lowest 
possible radial component.    

Since in the interior of the inner light surface $\Omega<\Omega_z$ and $f>0$ 
we conclude that $\eta_+$ is negative there (notice that $H_\phi<0$) and, 
thus, charged particles are bound to slide towards the black hole. 
As the result, no magnetic field 
line within the inner light surface can turn away from the black hole 
unless this takes place in the region where the force-free approximation 
breakes down (e.g. within the equatorial current sheet of the 
magnetospheric Wald problem, see figure \ref{wal-com}.)    

Similarly, one finds that in the outer superluminal region $\eta_-$ 
is positive and, thus, charged particles are forced to slide away from 
the hole (Basically the same analysis was used by Goldreich \& Julian (1969) 
with application to pulsar winds.) 

\subsection{Light surfaces and Alfv\'en critical surfaces}

The normal wavespeeds of Alfv\'en waves in force-free electrodynamics 
in FIDO's frame are given by

\begin{equation}
   \mu_{\pm} = \bmath{\mu}_\pm \cdot \bmath{n}
\label{d1}
\end{equation}

\noindent
where 

\begin{equation}
    \bmath{\mu}_\pm = \left(
        \bD\times\bB \pm \bB\sqrt{{B}^2-{D}^2}
               \right)/B^2
\label{d2}
\end{equation} 
\cite{Kom02a}.
By definition, at a critical Alfv\'en surface with the outgoing unit normal $\bn$   
either $\mu_+$ or $\mu_-$ changes its sign. As $\bmu_\pm$ are given by exactly 
the same expressions as $\bv_\pm$ (\ref{f2}) we immediately obtain that  

\begin{equation}
    \mu_\pm = (\spr{n}{B}_p) \eta_\pm. 
\label{d3}
\end{equation}
Thus, in the force-free limit the Alfv\'en surfaces coincide with 
the light surfaces. If $\bB_p$ is outgoing then $\mu_+$ corresponds 
to the outgoing wave and vanishes at the inner light surface. Inside
this surface $\mu_+<0$ and, thus, no Alfv\'en wave generated in this 
region can escape from it. Similarly, any Alfv\'en wave generated in 
the exterior of the outer critical surface is bound to remain there.  

Koide (2003) claimed that in the force-free limit the inner Alfv\'en 
surface coincides with the event horizon. This is obviously incorrect 
due to the property (v) of Appendix A1.  

\section{Boyer-Lindquist and Kerr-Schild coordinates}

In the Boyer-Lindquist coordinates $\{t,\phi,r,\theta\}$ 
the Kerr metric is 

\begin{equation}
   ds^2=g_{tt}dt^2+2g_{t\phi}dtd\phi+g_{\phi\phi}d\phi^2 +
        g_{rr}dr^2 + g_{\theta\theta}d\theta^2, 
\label{BLM}
\end{equation}
where 
$$ 
g_{tt} = z-1, \quad g_{t\phi}= -za\sin^2\!\theta, \quad 
g_{\phi\phi}=\Sigma\sin^2\!\theta/\rho^2, \quad
$$
$$
g_{rr}=\rho^2/\Delta,\quad g_{\theta\theta}=\rho^2,
$$
and 
\begin{eqnarray} 
  \nonumber
  \rho^2 &=& r^2 +a^2\cos^2\!\theta, \\ 
  \nonumber
   z &=& 2r/\rho^2,\\
  \nonumber
   \Sigma &=& (r^2+a^2)^2-a^2\Delta\sin^2\!\theta, \\
  \nonumber
   \Delta &=& r^2+a^2-2r. 
\end{eqnarray}
Notice that we use such units where the black hole mass $M=1$. 
At the event horizon, where $\Delta=0$, one has $g_{rr}=\infty$ which 
is the well know coordinate singularity of the Boyer-Lindquist
coordinates. The shift vector is 
   
\begin{equation}
    \bbeta = (g_{t\phi}/g_{\phi\phi},0,0),  
\label{BLFIDO}
\end{equation}
so the Boyer Lindquist FIDO is in the state of purely azimuthal 
motion with constant angular velocity. From (\ref{ncov}) one finds that 
$n_\phi=0$ and, thus, FIDO is also ZAMO (Zero Angular Momentum Observer, 
Bardeen at al.,1973.)  
At the even horizon its world line turns space-like and this is another
consequence of the coordinate singularity. 
This coordinate singularity can be removed via the following integrable 
coordinate transformation, also singular at the event horizon:
\begin{eqnarray} 
  \nonumber
  dt&\rightarrow&dt+G(r)dr, \\
  \nonumber
  d\phi&\rightarrow&d\phi+H(r)dr,  
\end{eqnarray}
where 
$$
   G(r)=-2r/\Delta, \quad H(r)= -a/\Delta.
$$
(Since the transformation law for $t$ depends on $r$ this introduces a 
different foliation of space-time.) The corresponding transformation matrices
are 
\begin{equation} 
{\cal A}^{\mu'}_\mu = \left( 
   \begin{array}{cccc} 
     1 & 0 & G & 0 \\
     0 & 1 & H & 0 \\ 
     0 & 0 & 1 & 0 \\ 
     0 & 0 & 0 & 1  
   \end{array} \right) ,
\end{equation}
\begin{equation} 
({\cal A}^{-1})^{\mu}_{\mu'} = \left( 
   \begin{array}{cccc} 
     1 & 0 & -G & 0 \\
     0 & 1 & -H & 0 \\ 
     0 & 0 & 1 & 0 \\ 
     0 & 0 & 0 & 1  
   \end{array} \right).
\end{equation}
Here we assume that 1) the primed indexes refer to the Boyer-Lindquist 
coordinates; 2) the upper index indicates the matrix row and the lower 
index indicates the matrix column. This allows us to apply the Einstein summation
rule to the transformation laws, e.g. 
$$
   dx^{\mu'}= {\cal A}^{\mu'}_\mu dx^\mu.
$$

In the new coordinates, known as the Kerr-Schild coordinates, the Kerr 
metric is 
\begin{equation}
   \begin{array}{rl}
   ds^2= & g_{tt}dt^2+2g_{t\phi}dtd\phi+2g_{tr}dtdr+\\
        & g_{\phi\phi}d\phi^2 + 2g_{r\phi}d\phi dr +
        g_{rr}dr^2 + g_{\theta\theta}d\theta^2, 
   \end{array}
\label{KSM}
\end{equation}  
where
$$
  g_{tt} = z-1, \quad g_{t\phi}= -za\sin^2\!\theta, \quad
  g_{tr} = z,
$$
$$
  g_{\phi\phi}= \Sigma \sin^2\theta/\rho^2,\quad
  g_{r\phi}=-a\sin^2\theta(1+z), \quad
$$
$$
  g_{rr}= 1+z, \quad
  g_{\theta\theta}=\rho^2. 
$$
Thus the spatial metric tensor is  

\begin{equation} 
  \gamma_{ij} = \left(\begin{array}{ccc} 
       \Sigma \sin^2(\theta)/\rho^2 & -a\sin^2(\theta)(1+z) & 0 \\
       -a\sin^2(\theta)(1+z) & 1+z & 0 \\
        0 & 0 & \rho^2 
       \end{array} \right).  
\label{gamma}
\end{equation}
As $g_{r\phi} \not= 0$ the spatial coordinates are no longer orthogonal.
However, at infinity this tensor has exactly
the same form as the metric tensor of Euclidean space in spherical
coordinates. This is no surprise as for $r \to \infty$ one has
$F(r),H(r)\to 0$ and the Kerr-Schild coordinates become identical to
the Boyer-Lindquist coordinates. 
   
The Kerr-Schild lapse function and its shift vector are 
 
\begin{eqnarray} 
    \alpha &=& 1/\sqrt{1+z}, \\
    \beta^i &=& (0,\frac{z}{1+z},0).
\end{eqnarray}
Thus, the Kerr-Schild FIDO is moving  radially towards the true
singularity at $r=0$. 

It is illuminating to determine the motion of the Kerr-Schild FIDO 
in the Boyer-Lindquist coordinates. Simple calculations give 
$$
 v^{\phi}=2ar/\Sigma, \quad v^{r}=-2r\Delta/\Sigma, 
 \quad v^{\theta}= 0.
$$ 
Thus, the Kerr-Schild FIDO has the same angular velocity as 
the Boyer-Lindquist FIDO but also moves radially towards 
the true singularity. 

The tensor transformation laws 
$$
   {\cal U}_\mu = {\cal A}_\mu^{\mu'} {\cal U}_{\mu'}
$$
and
$$  
   F_{\mu\nu} = {\cal A}_\mu^{\mu'}{\cal A}_\nu^{\nu'} F_{\mu'\nu'}
$$ 
allow us to relate the components of the four-potential and the 
electromagnetic field tensor in both coordinate systems. 
It is easy to see that all components of the four potential except 
${\cal U}_r$ and all components of $F_{\mu\nu}$ with $\mu,\nu \not= r$ 
are invariant. This immediately shows that the scalar potential 
$\Phi=-{\cal U}_t$, the magnetic flux function $\psi=2\pi{\cal U}_\phi$,
$H_\phi =\Fs_{t\phi}$ and $E_\phi =F_{\phi t}$ are the same in both 
coordinate systems. Since the angular velocity of magnetic field is given 
by 
$$
   \Omega = - F_{t\theta}/F_{\phi\theta},
$$
it is also invariant.

\section{Numerical method}

\subsection{Augmented system of the method of generalized 
Lagrange multiplier}

From eq.(\ref{Faraday}) one finds 
\begin{equation} 
  \partial_t(\vdiv{B}) = 0.  
\label{inforcing}
\end{equation} 
This well known result shows that it is sufficient to enforce 
the divergence free condition (\ref{divB}) only for the 
initial solution and it will automatically be satisfied at any 
$t$.  Unfortunately, straightforward applications of many numerical 
schemes, perfectly suitable for other hyperbolic systems of 
conservation laws, fail to deliver a good result in the case of electrodynamics 
and MHD simply because their discrete equations are not consistent with 
any discrete analogue of (\ref{inforcing}). 
In particular, this applies to the method of Godunov 
which has many beneficial properties and is currently considered 
as generally superior to many other numerical schemes for hyperbolic 
systems. There have been many attempts to find a cure for this 
``div-B problem'', though no perfect solution seems to have been 
found so far ( see the review in Dedner et al.,2002.)        

One of the ways to handle this problem involves construction of a 
somewhat different system of differential equations, the ``augmented 
system'', where the divergence free condition (\ref{divB}) is 
no longer included and $\vdiv{B}$ may be transported and/or 
dissipated like other dynamical variables. The idea is 
not to enforce the divergence free condition exactly but to 
promote a natural evolution of the system towards a divergence 
free state. Provided the augmented system is hyperbolic, it can 
be solved numerically using the Godunov method \cite{Mun,Ded}.     

Following this idea we propose to modify eq.(\ref{Maxw1}) as follows

\begin{equation} 
  \nabla_\beta  \Fs^{\alpha \beta} +\nabla^\alpha \Psi 
  + \kappa k^\alpha \Psi = 0, 
\label{Maxw1a}
\end{equation} 
where $\Psi$ is a scalar field which we shell call 
``the pseudo-potential'' and  $\kappa=$const. As before   
$k^\nu = \partial_t$, where $t$ is the global time coordinate  
of space-time.  
From equation (\ref{Maxw1a}) one immediately obtains the evolution 
equation for $\Psi$ 

\begin{equation} 
  \Box\Psi+ \kappa \nabla_t \Psi =0,  
\label{Del-Psi}
\end{equation} 
where $\Box = \nabla_\nu \nabla^\nu$ is the d'Alembert operator. 
This is the well known ``telegraph equation'' and it prescribes 
both propagation and dissipation of $\Psi$.  In fact, $\vdiv{B}$ 
is governed by a similar equation. Indeed, from the time component  
of eq.(\ref{Maxw1a}) one obtains  
$$
  \nabla^t \Psi + \kappa \Psi = \frac{1}{\alpha} \vdiv{B}, 
$$
whereas from the spatial part one has 
$$
  \nabla_t\left( \frac{\vdiv{B}}{\alpha} \right) = 
         - \nabla_i \nabla^i \Psi. 
$$
Combining, these two equations one finds first 

$$ 
   \nabla^t\nabla_t \left(\frac{\vdiv{B}}{\alpha} \right) + 
   \kappa \nabla_t \left( \frac{\vdiv{B}}{\alpha} \right) + 
   \nabla_i\nabla^i(\nabla^t \Psi + \kappa \Psi) = 0
$$ 
and, finally,   

\begin{equation} 
  \Box\left( \frac{\vdiv{B}}{\alpha} \right ) + 
  \kappa \nabla_t\left( \frac{\vdiv{B}}{\alpha} \right) =0,  
\label{Del-divb}
\end{equation} 

Direct 3+1 splitting of eq.(\ref{Maxw1a}) gives us 
two conservation laws:  

\begin{equation} 
\begin{array}{l}
   \Pd{t}\left(\sqrt{\gamma} \frac{\Psi}{\alpha} \right) + 
     \Pd{j}\left(\sqrt{\gamma}  (B^j-\frac{\Psi}{\alpha}\beta^j)\right) = \\
     \qquad\qquad\qquad\qquad  \Psi\left( \kappa\alpha\sqrt{\gamma} -
     \Pd{j} ( \sqrt{\gamma}\frac{\beta^j}{\alpha}) \right)  
\end{array}
\label{Psi}
\end{equation} 
and

\begin{equation} 
\begin{array}{l}
  \Pd{t}\left(\sqrt{\gamma} (B^i+\frac{\Psi}{\alpha}\beta^i) \right) + 
  \Pd{j} \left(\sqrt{\gamma}(e^{ijk}E_k +\alpha\Psi g^{ij})  \right) = \\ 
  \qquad \qquad \qquad \qquad \qquad \qquad \qquad 
   \Psi \Pd{j}(\alpha\sqrt{\gamma} g^{ij}).  
\end{array}
\label{Far3}
\end{equation}
Summarizing, we have constructed a system of two vector conservation 
laws, equations (\ref{Amp1},\ref{Far3}), 
and one scalar conservation law, equation (\ref{Psi}).

\subsection{Numerical scheme}

All 7 evolution equations of the augmented system,  
as well as the equations of the general relativistic 
electrodynamics itself, are conservation laws. 
We can write them as a single abstract vector equation 
of the form    

\begin{equation} 
  \Pd{t} ({\sqrt{\gamma}\cal Q}^K) + 
         \Pd{j}( {\sqrt{\gamma}\cal F}^{Kj}) = 
            \sqrt{\gamma} ({\cal S}_{(g)}^K + {\cal S}_{(d)}^K),     
\label{CONS}
\end{equation}
where 

\begin{equation}
  {\cal Q}^K = \left( 
     \frac{\Psi}{\alpha}, 
     B^i+\frac{\Psi}{\alpha}\beta^i, 
     D^i \right) 
\end{equation}
are the conserved variables, where $i$ is the index of spatial coordinates, 

\begin{equation}
   {\cal F}^{Kj} =\left( 
     B^j -\frac{\Psi}{\alpha}\beta^j, 
      e^{ijk}E_k+\alpha\Psi g^{ij}, 
      - e^{ijk}H_k \right)
\end{equation}
are the corresponding hyperbolic fluxes,

\begin{equation}
{\cal S}_{(g)}^K = 
\left(-\frac{\Psi}{\sqrt{\gamma}}\Pd{k}(\sqrt{\gamma}\frac{\beta^k}{\alpha}), 
        \frac{\Psi}{\sqrt{\gamma}} \Pd{k}(\sqrt{\gamma}\alpha g^{ik}), 
        \rho( -\alpha v_d^i + \beta^i) \right)
\end{equation}
are the geometrical source terms, and 

\begin{equation}
{\cal S}^K_{(d)} = \left( 
       \alpha \kappa\Psi, 0^i, 
       -\alpha(\sigma_{\parallel} D_{\parallel}^i + 
       \sigma_{\perp} D_{\perp}^i) \right)  
\end{equation}
are the potentially stiff dissipative source terms.

To handle stiff source terms we use the time-step splitting 
technique \cite{Lev}. Namely, each time-step we first integrate 
the truncated system 

\begin{equation} 
  \Pd{t} ({\sqrt{\gamma}\cal Q}^K) + 
         \Pd{j}( {\sqrt{\gamma}\cal F}^{Kj}) = 
            \sqrt{\gamma} {\cal S}_{(g)}^K ,     
\label{S1}
\end{equation} 
The result is then considered as the initial solution for another 
truncated system  

\begin{equation} 
  \Pd{t} {\cal Q}^K = 
            {\cal S}_{(d)}^K ,     
\label{S2}
\end{equation} 
which is integrated over the same time step.   

In order to integrate (\ref{S1}) we apply the Godunov method, 
slightly modified to accommodate for the space-time curvature 
\cite{Pons}.  Integrating (\ref{S1}) over a spacetime cell 
$(\Delta t\times \Delta x^1 \times \Delta x^2 \times \Delta x^3)$ one obtains

\begin{equation} 
  {\cal Q}^K(t+\Delta t) =  {\cal Q}^K(t) 
    -\frac{\Delta t}{\Delta V} \sum\limits_j {\cal F}^{Kj} \Delta S_j   
    +{\cal S}_{(g)}^K \Delta t,     
\label{SCH}
\end{equation} 
where the summation is taken over all spatial faces of the cell. 
In this equation ${\cal Q}$ is the vector of conserved variables 
averaged over the cell volume, ${\cal S}_{(g)}$ is the vector 
of source terms averaged over the cell volume and the time interval, 
and ${\cal F}^{Kj}$ are averaged over the corresponding j-face of 
the cell and the time interval. If the computational cells are created 
by coordinate surfaces, as it is in our case, then 
$\Delta V = \sqrt{\gamma} \Delta x^1 \Delta x^2 \Delta x^3$ 
and   
$\Delta S_i = \pm \sqrt{\gamma} \Delta x^j \Delta x^k$, where 
$i\not=j\not=k$. 

Equation (\ref{SCH}) allows us to advance the numerical solution 
in time provided ${\cal F}^{Kj}$ and ${\cal S}_{(g)}$ are known. 
In the Godunov method, the fluxes ${\cal F}^{Kj}$ are found via solving 
the corresponding Riemann problem at the cell interfaces  
\cite{Lev}. Following Pons et al.\shortcite{Pons} we setup such problems   
utilizing the ``primitive'' solution vector, $\bP$, which includes  
the components $\check{E}^{\hat{i}}=D^{\hat{i}}$ and  $B^{\hat{i}}$ of 
electric and  magnetic field as measured in the orthonormal basis 
of local FIDO, $\bP^K=(\Psi, B^{\hat{i}}, D^{\hat{i}})^t$. 
This allows us to consider the Riemann problem as observed 
in the locally inertial frame of the FIDO initially placed at
the cell interface and use a special relativistic Riemann solver 
to find its solution in this frame. At this stage, it is important 
to take into account the finite speed of the interface relative to FIDO 
, $\beta^{\hat{i}}$ \cite{Pons}.    
  
In the locally inertial frame with the Riemann discontinuity 
normal to the x-axis the relevant system of 1D equations 
is linear 

\begin{equation} 
   \Pd{t} \bP^K + {\cal A}^K_L \Pd{x}\bP^L = 0 
\label{1Dsystem}
\end{equation}
with the Jacobean matrix 
\[
{\cal A} =
\left[
\begin{array}{ccccccc}
 0 & 1 & 0& 0 & 0 & 0 & 0 \\
 1 & 0 & 0& 0 & 0 & 0 & 0 \\
 0 & 0 & 0& 0 & 0 & 0 & -1 \\
 0 & 0 & 0& 0 & 0 & 1 & 0 \\
 0 & 0 & 0& 0 & 0 & 0 & 0 \\
 0 & 0 & 0& 1 & 0 & 0 & 0 \\
 0 & 0 & -1& 0 & 0 & 0 & 0 \\
\end{array}
\right] .
\]
The eigenvalues of this matrix 
$$
  \mu_{(1)} = \mu_{(2)} = \mu_{(3)} =1; 
$$
$$
\mu_{(4)}=0; 
$$
$$
  \mu_{(5)}=\mu_{(6)}=\mu_{(7)} =-1  
$$
provide the wavespeeds of the hyperbolic waves. Other properties
of these waves are given by the eigenvectors of ${\cal A}$. 
The right eigenvectors are    

\begin{equation} 
\begin{array}{ccc}  
\br_{(1)} & = & (1,1,0,0,0,0,0) \\
\br_{(2)} & = & (0,0,-1,0,0,0,1) \\
\br_{(3)} & = & (0,0,0,1,0,1,0) \\
\br_{(4)} & = & (0,0,0,0,1,0,0) \\
\br_{(5)} & = & (-1,1,0,0,0,0,0) \\
\br_{(6)} & = & (0,0,1,0,0,0,1) \\
\br_{(7)} & = & (0,0,0,-1,0,1,0) \\
\end{array}
\end{equation}
Since ${\cal A}$ is symmetric, its left eigenvectors, $l^K_{(i)}$, 
coincide with the corresponding right eigenvectors: 
\begin{equation} 
 l^K_{(i)} = r^K_{(i)}, 
\end{equation}

It is easy to see that the solutions 2,3,6, and 7 describe  
the usual electromagnetic waves. The solution 4 simply reflects 
the fact that all waves of vacuum electrodynamics are 
transverse and any discontinuity in the normal component of 
electric field is due to a surface electric charge distribution.   
the solutions 1 and 5 describe new waves which do not exist in 
electrodynamics; they transport $\Psi$ and $\vdiv{B}$ with the 
speed of light.     

The solution to the Riemann problem with the left and the right 
states, $\bP_{(l)}$ and $\bP_{(r)}$, and the interface 
speed $\beta^x$ is

\begin{equation} 
  \bP = \left\{ 
\begin{array}{ccc} 
            \bP_{(l)} & \text{if} & \beta^x < -1; \\           
            \bP_{(r)} & \text{if} & \beta^x > +1; \\ 
            \bP_{(l)} + \sum\limits_{i=1,3} 
                \kappa_{(i)} \br_{(i)} & \text{if} &  -1 < \beta^x <+1, \\
\end{array}
            \right. 
\end{equation}
where 

$$  
  \kappa_{(i)} = \frac{(\bP_{(r)}-\bP_{(l)} )\cdot \br_{(i)}}
                  {\br_{(i)}\cdot \br_{(i)}}.
$$ 
Notice, that the 4th wave is ignored and the x-component of $\bD$ 
is set to be the mean value of the left and the right states 

\begin{equation} 
 D^{\hat{x}} = 0.5 (D^{\hat{x}}_{(l)} + D^{\hat{x}}_{(r)}) .  
\end{equation}

In order to make the scheme second order in space and time we 
introduce slope-limited linear distribution of primitive variables within 
each cell and use the half-time-step solution following the description  
in \cite{Falle,Kom99}.

\subsection{Test simulations}  

To test the code we carried out a number of test simulations. Here we
describe only two of them. In both cases the space-time metric is Minkowskian, 
so $\bE=\bD$ and $\bH=\bB$.   

The dissipative time scales of our resistivity model 
is tied to the computational time-step, $\Delta t$. Namely, we set

\begin{equation} 
   \sigma_{\parallel} = d/\Delta t, 
\label{sigma_perp}
\end{equation}
and, following the analysis of Sec.2.3, we make the cross field  resistivity 
strongly dependent on $B^2-D^2$: 

\begin{equation} 
   \sigma_{\perp} = \sigma_\parallel \left\{ \begin{array}{rl} 
       0 & \text{if} B^2 \ge D^2 \\  
       b (D_{\perp}-D_{\perp}^*)/D_{\perp}^*  
        & \text{if} B^2 < D^2
       \end{array}, \right.  
\label{sigma_para}
\end{equation}
where $(D_{\perp}^*)^2 = B^2-D^2_{\parallel}$. This model provides continuous 
$\sigma_\perp$ at $D^2=B^2$ which is preferable for numerical reasons.   
The corresponding solutions to equations (\ref{S2}) are 

\begin{equation} 
    D_\parallel(t) = D_\parallel(0) e^{-\sigma_\parallel t},  
\label{E_para}
\end{equation}
and

\begin{equation} 
    D_\perp(t)=D_\perp^* + 
    \frac{D_\perp^*(D_\perp(0)-D_\perp^*) e^{-b\sigma_\parallel t} }
    {D_\perp(0)+(D_\perp^*-D_\perp(0)) e^{-b\sigma_\parallel t} } .
\label{E_perp}
\end{equation}
In fact, each time-step we first evolve $D_\parallel$ as in eq.(\ref{E_para}) 
and then $D_\perp$ according to eq.(\ref{E_perp}). In most of the problems 
we use $d\le 1$ and $b=0.1$.  
In addition, we assume that the drift current is described by 
eq.(\ref{jdrift2}). 
Although this model resistivity is too high compared to the physical one, 
any much lower model resistivity would not allow to resolve current sheets 
(and would become lower that the numerical resistivity.)

\begin{figure}
\leavevmode
\epsffile[0 0 216 216]{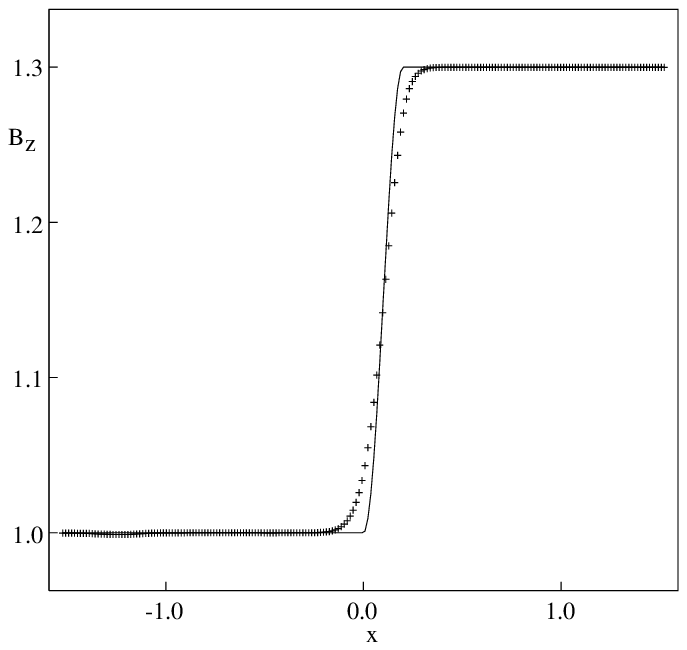}
\caption{Stationary Alfv\'en wave. The exact solution and the initial 
numerical solution are shown by the continuous line. The crosses show 
the numerical solution at $t=2$}   
\label{aw-test}
\end{figure}

\begin{figure*}
\leavevmode
\epsffile[0 0 480 216]{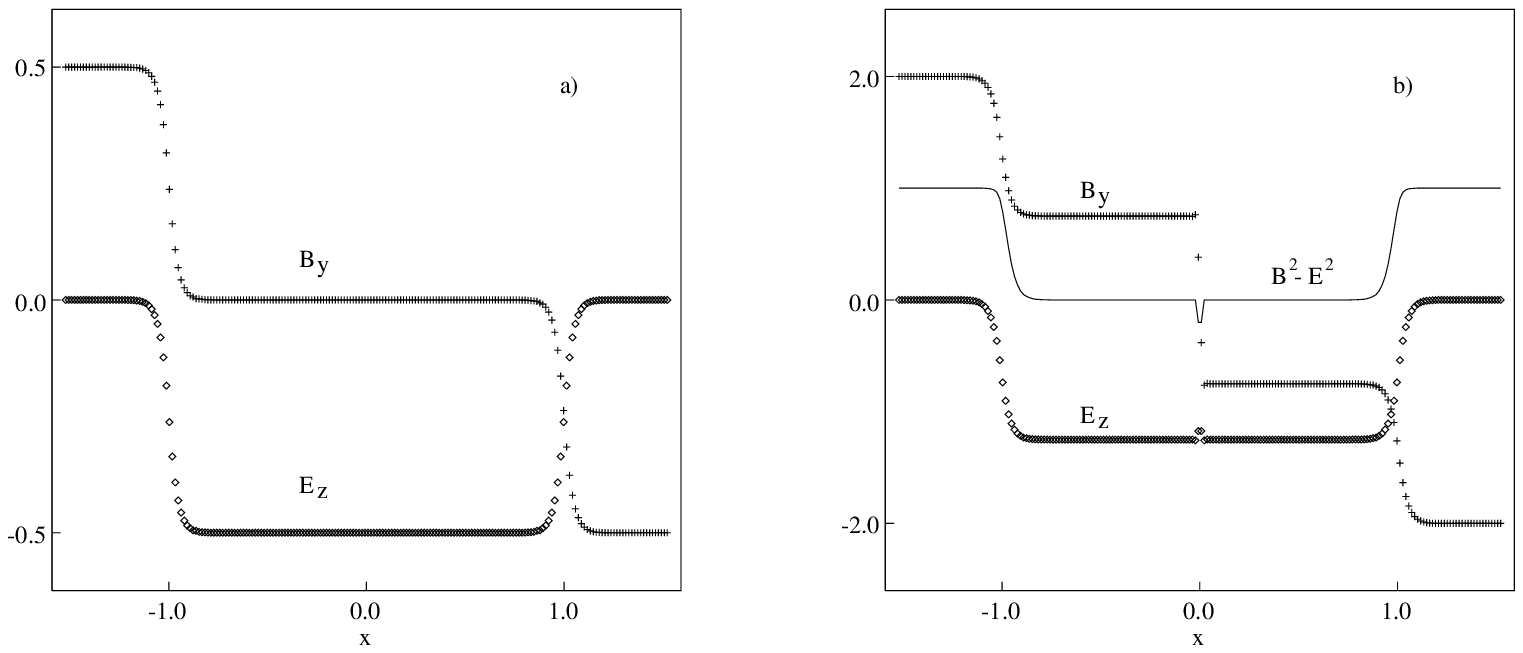}
\caption{Riemann problem described in Section C3.2. 
{\it Left panel:} Numerical solution at $t=1$ for 
$B_0=0.5$.
{\it Right panel:} Numerical solution at $t=1$ for 
$B_0=2.0$. The current sheet is located at $x=0$. }
\label{sod}
\end{figure*}

\subsubsection{Alfv\'en wave} 

The system of force-free electrodynamics allows two types of hyperbolic 
waves: 1) the fast waves, which have the same properties as linearly 
polarized waves of vacuum electrodynamics, and 2) the Alfv\'en waves, 
which are similar to the Alfv\'en waves of relativistic MHD \cite{Kom02a}.
The analytical solution for the Alfv\'en wave used in this test 
describes a stationary wave with $B_x=B_y=D_z=1$, $D_y=0$, 

$$ 
   B_z(x)=\left\{ 
   \begin{array}{rl} 
    1 & \text{for} x < 0; \\ 
    1 + 0.15(1+\sin{5\pi(x-0.1)})& \text{for} 0 < x < 0.2; \\ 
    1.3 & \text{for} x > 0.2;  
   \end{array} \right.
$$ 
and 
$$
    D_x =-B_z . 
$$ 
The computational grid included 200 identical cells equally spaced 
between $x=-1.5$ and $x=1.5$.

Figure \ref{aw-test} shows the solution at $t=1$. While an electromagnetic wave 
would cover a distance $\Delta x =1$ during this time, the Alfv\'en wave 
remains stationary and only spreads due to finite resistivity.

\subsubsection{Current sheet} 

Here we consider the following Riemann problem: 
$$ 
    \bD = 0, \text{ } B_z=0, \text{ } B_x=1 
$$  
$$ 
  B_y=\left\{ 
   \begin{array}{rl}  
     B_0 & \text{for} x<0;\\ 
     -B_0 & \text{for} x>0; 
   \end{array} \right.
$$
The symmetry of this problem implies $B_y=0$ at the interface 
at time $t>0$. If $B_0<1$ then this problem has a force-free solution 
describing two fast waves, switching-off the tangential component 
of magnetic field and switching-on the z-component of electric field
 (up to the value of $-B_0$.) 
In fact, the vacuum electrodynamics has exactly the same solution to this 
problem.  The numerical solution at $t=1$ for $B_0=0.5$ is shown in the panel 
a) of figure \ref{sod}. The computational grid is uniform with 200 cells.  

If $B_0>1$ then the vacuum electrodynamics still allows solutions of 
this kind. However, the force-free electrodynamics has no solution 
in this case because $B^2-D^2$ vanishes when  $|B_y|$ drops down to the value of  
$(B_0^2+1)/2B_0$. At this point the fast wave is terminated by 
increased cross-field conductivity that locks $B^2-D^2$ to 0. Downstream of 
these waves, the Poynting flux is directed towards the interface located at $x=0$ 
where the electromagnetic 
energy dissipates in a current sheet. The panel b) of figure \ref{sod} 
shows the numerical solution for $B_0=2$ at time $t=1$ on a uniform 
grid with 200 cells.  As expected, the global solution is not sensitive 
to the exact value of the cross field conductivity which only effect the 
structure of the current sheet, the higher $\tau_\perp$ leading to the wider
current sheet.

\end{document}